\documentclass[a4paper, 12pt]{article}
\usepackage{a4}
\usepackage{amsfonts}
\usepackage{amssymb}
\usepackage{cite}
\usepackage{epsfig}
\usepackage{psfig}
\usepackage{amsmath}
\usepackage{cite}
\usepackage{epsfig}
\usepackage{psfig}

\DeclareMathOperator{\ch}{ch}

\DeclareMathOperator{\dP}{dP}

\author{}

\newcommand{\be}{\begin{equation}}
\newcommand{\ee}{\end{equation}}
\newcommand{\bea}{\begin{eqnarray}}
\newcommand{\eea}{\end{eqnarray}}
\newcommand{\mbb}{\mathbb}

\newcommand{\1}{{\bf 1}}

\newcommand{\4}{{\bf 4}}
\newcommand{\5}{{\bf 5}}
\newcommand{\6}{{\bf 6}}

\newcommand{\ov}{\overline}
\def\IR{\relax{\rm I\kern-.18em R}}

\def\IP{\relax{\rm I\kern-.18em P}}
\def\inbar{\vrule height1.5ex width.4pt depth0pt}
\def\IC{\relax\,\hbox{$\inbar\kern-.3em{\rm C}$}}

\def\K3{{\bf K3}}

\def\ov{\overline}
\def\preal{{\rm Re\,}}

\begin{document}

\title{
\begin{flushright} \vspace{-2cm}
{\small MPP-2006-16 \\
 \small LMU-ASC 11/06\\
\vspace{-0.35cm}
hep-th/0603015} \end{flushright}
\vspace{4.0cm}
Heterotic GUT and Standard Model Vacua from simply connected Calabi-Yau Manifolds
\quad
}
\vspace{1.0cm}
\author{\small Ralph~Blumenhagen$^{\heartsuit}$, Sebastian Moster$^{\heartsuit}$ and  Timo Weigand$^{\heartsuit,\spadesuit}$}

\date{}

\maketitle

\begin{center}
\emph{$^{\heartsuit}$ Max-Planck-Institut f\"ur Physik, F\"ohringer Ring 6, \\
  80805 M\"unchen, Germany } \\
\vspace{0.1cm}
\emph{$^{\spadesuit}$ Arnold-Sommerfeld-Center for Theoretical Physics, Department f\"ur Physik,
Ludwig-Maximilians-Universit\"at  M\"unchen, Theresienstra\ss e 37, 80333 M\"unchen, Germany}\\
\vspace{0.2cm}

\tt{blumenha,moster,weigand @mppmu.mpg.de}
\vspace{1.0cm}
\end{center}
\vspace{1.0cm}

\begin{abstract}
\noindent  
We consider four-dimensional supersymmetric compactifications of the $E_8\times E_8$
heterotic string on Calabi-Yau manifolds endowed
with vector bundles with structure group $SU(N)\times U(1)$
and five-branes. 
After evaluating the Green-Schwarz mechanism and  deriving the generalized
Donaldson-Uhlenbeck-Yau condition including the five-brane moduli,
we show that this construction can give rise to GUT models
containing $U(1)$ factors like flipped $SU(5)$ or directly the Standard Model even on simply connected Calabi-Yau manifolds.
Concrete realizations of three-generation models on elliptically fibered
Calabi-Yau manifolds are presented. They exhibit the most attractive features of flipped $SU(5)$ models such as doublet-triplet splitting and proton stability. In contrast to conventional GUT 
string models, the tree level relations among the Standard Model
gauge couplings at the GUT scale are changed.

\end{abstract}

\thispagestyle{empty}
\clearpage

\tableofcontents

\section{Introduction and Summary}

If nature really knows about superstring theory, it must of course be possible to embed the observed Standard Model of particle physics into the theory. Most attempts to verify this hope explicitly and thereby  provide the first ultra-violet completion of the low-energy effective quantum field theory initially focused on the ${\mbox{}E_8 \times E_8}$ heterotic corner of the M-theory star. The attitude has been to break the original ten-dimensional gauge group down to phenomenologically appealing GUT groups and to recover the observed light matter as irreducible representations of the latter \cite{Witten:1985bz}. The classic approaches comprise compactifications on toroidal orbifolds or more generally on Calabi-Yau manifolds endowed with stable holomorphic vector bundles. In the latter case, giving VEVs to the internal field strengths with values in, say, an $SU(4)$ or $SU(5)$ subgroup of one of the $E_8$ factors leads to the respective commutant in $E_8$,  $SO(10)$ or $SU(5)$ as the four-dimensional gauge group. The ${\bf 248}$ representation of $E_8$ very naturally splits into the respective GUT multiplets which incorporate the chiral fermions of the Standard Model. However, in this construction it is impossible to realize the GUT breaking further down to $SU(3) \times SU(2) \times U(1)_Y$ via a field theoretic Higgs mechanism, simply because the required vector-like pairs from which the GUT Higgs could arise are not present in the particle spectrum\footnote{Note, however, that in the context of higher-level Kac-Moody algebras GUT Higgses {\emph {can}} be realized.}. This necessitates the use of discrete Wilson lines as an alternative GUT breaking mechanism and considerably complicates the construction of heterotic Standard Model vacua. The point is that in order to have these Wilson lines at our disposal, we need to work on Calabi-Yau manifolds with non-zero first fundamental group, and constructing them explicitly takes quite some effort. It has been one of the recent triumphs of string model building to provide classes of such Calabi-Yau manifolds as quotients of manifolds under an appropriate freely-acting orbifold group and to construct non-abelian vector bundles on them \cite{Donagi:2000zf,Donagi:2000zs,Donagi:2000fw,Braun:2004xv,Andreas:1999ty,Curio:2004pf}. Globally defined realistic models from $SU(5)$ GUT in this context have been provided in \cite{Bouchard:2005ag}. For realistic models from $SO(10)$ see \cite{Braun:2005ux,Braun:2005nv}.\footnote{For these latter constructions, stability of the visible sector bundle has been proven in \cite{Gomez:2005ii,Braun:2006ae}.  However, as of this writing, no stable hidden bundle has been found in order to satisfy the tadpole cancellation condition. As they stand, the models are therefore not yet globally defined.} A recent construction of realistic models in the setup of heterotic toroidal orbifolds can be found in \cite{Buchmuller:2005jr}.

Independently of the heterotic model building industry, the discovery of D-branes has opened up a complementary path to incorporating gauge interactions into string theory, more precisely the Type II theory or orientifolds thereof. The extensive analysis of intersecting D-brane models  in Type IIA (for the most recent review and references consult \cite{Blumenhagen:2005mu}) demonstrated that not only non-abelian, but also abelian background gauge instantons can be of phenomenological interest. The connection between the two pictures is that the objects mirror dual to D6-branes at angles are magnetized D9-branes in Type I theory, which in turn are S-dual to abelian background bundles in the $SO(32)$ heterotic theory. It is therefore of obvious  relevance to explore the usually neglected use of non-trivial line bundles\footnote{For some early references see \cite{Green:1984bx,Distler:1987ee,Aldazabal:1996du,Lukas:1999nh} and more recently \cite{Andreas:2004ja}.} in heterotic compactifications with the hope of extending our model building possibilities beyond the classic embedding of vector bundles with vanishing first class only. Likewise, one might wonder if turning on also non-abelian gauge bundles on D9-branes in Type I might lead to interesting constructions. A step into that direction has been performed in \cite{Blumenhagen:2005pm,Blumenhagen:2005zg,Blumenhagen:2005zh}, where it has been demonstrated that vacua with Standard-Model like gauge group and matter do indeed exist in the framework of these non-abelian braneworlds.

Building upon and extending the results of \cite{Blumenhagen:2005ga,Blumenhagen:2005zg}, it is the aim of this article to construct realistic vacua from the $E_8 \times E_8$ string with general (non-)abelian vector bundles. We will see that the embedding of $SU(4) \times U(1)$ bundles very naturally leads to the breaking of $E_8$ down to the GUT group $SU(5) \times U(1)_X$. In fact, the quantum numbers of  the resulting matter representations are precisely those of the flipped SU(5) model \cite{Barr:1981qv,Derendinger:1983aj}. Previous attempts to realize this scenario from different string constructions include \cite{Antoniadis:1989zy,Chen:2005cf}. A priori, the $U(1)_X$ receives a Stueckelberg mass due to the Green-Schwarz mechanism so that we seem to be back in the classic Georgi-Glashow scenario and face the problem pointed out above that we need Wilson lines for GUT breaking. By contrast, we can alternatively embed the same line bundle also into the second $E_8$ in such a way that the $U(1)_X$ is now a linear combination of the two abelian factors and remains massless. Remarkably, due to the particular way how the MSSM quantum numbers are assigned to the various representations in flipped $SU(5)$, the GUT Higgs can now arise from a ${\bf 10}-{\bf \ov {10}}$ pair present in the spectrum. We can thus circumvent the necessity of working on Calabi-Yau manifolds with $\pi_1(CY_3) \neq 0$, thereby bringing a large number of geometric backgrounds back into the heterotic model building business. Following the same rationale, it is also possible to directly arrive at the MSSM gauge group from $E_8 \times E_8$ by employing $SU(5) \times U(1)$ bundles. The resulting matter representations carry exactly the MSSM quantum numbers.

 Our explicit model search has focused so far on elliptically fibered Calabi-Yau manifolds: Here we profit from the spectral cover construction as the working horse to arrive at stable holomorphic vector bundles \cite{Friedman:1997ih,Friedman:1997yq}. Besides, we will see that the elliptic fibration property of the Calabi-Yau admits a very natural solution to some of the constraints arising in our construction. Concretely, in both cases we embed the bundle $W=V \oplus L^{-1}$ into the first $E_8$ with $V$ being a $U(4)$ or $U(5)$ bundle, respectively, and the line bundle $L^{-1}$ chosen such that $c_1(W) = 0$. The bundle $V$ arises by first constructing a stable $SU(4)$ respectively $SU(5)$ bundle \`a la Friedman-Morgan-Witten and then twisting  it by a line bundle \cite{Blumenhagen:2005zg}. 
 Our search has provided a couple of models with precisely three chiral generations of MSSM matter and no further chiral exotics. The detailed analysis of the non-chiral part of the spectrum is of course essential and postponed to a future publication.    
 
At the technical level, the use of non-trivial line bundles makes a very careful investigation of anomalous $U(1)$s and the associated Green-Schwarz mechanism necessary. In this respect we will extend the analysis of \cite{Blumenhagen:2005ga} to vacua including also heterotic five-branes. Among the new features appearing is the presence of additional Green-Schwarz counter terms which have to cancel mixed abelian anomalies due to the presence of the five-branes. The existence of these additional terms is consistent with the recent analysis of six-dimensional compactifications of the heterotic string with line bundles and five-branes \cite{Honecker:2006dt}. They can be further justified by a direct derivation from heterotic M-theory. Even in the absence of abelian gauge factors, these terms contribute to the four-dimensional threshold corrections and lead to five-brane dependent expressions which have also been observed in reducing Witten's background solution to heterotic M-theory \cite{Lukas:1997fg,Lukas:1998hk}. Likewise, the one-loop correction of the Donaldson-Uhlenbeck-Yau equation of \cite{Blumenhagen:2005ga} receives five-brane dependent contributions. From the associated Fayet-Iliopoulos terms of the effective four-dimensional supergravity theory we will find a new D-term potential for the M5-brane in heterotic M-theory induced by abelian line bundles on either of the two ten-dimensional $E_8$-orbifold planes.

 Phenomenologically, our $SU(5) \times U(1)_X$ construction inherits the appealing features of the flipped $SU(5)$ scenario, in particular the natural solution of the doublet-triplet splitting problem and the automatic suppression of dangerous dimension-five operators leading to proton decay \cite{Antoniadis:1987dx}. In addition, as a consequence of the particular properties of our bundle embeddings, the Standard Model Higgs turns out to carry different quantum numbers than the lepton doublet, and thus  also dimension-four decay operators are absent. Since the massless $U(1)_X$ is actually a combination of two abelian factors from both $E_8$'s, the standard GUT relation for the tree-level MSSM gauge couplings is not satisfied; however, by taking into account also the threshold corrections which manifestly depend on the K\"ahler moduli of the internal manifold one can achieve the usual gauge coupling unification on a codimension-one hypersurface in the K\"ahler moduli space. After all, we should not forget that, like in all weakly coupled heterotic constructions, the four-dimensional Planck scale comes out too low and we should really promote the model to the strong coupling Horava-Witten regime \cite{Witten:1996mz}. This, too, is in fact nothing else than choosing part of the K\"ahler moduli of the model - in this case the size of the eleventh dimension - in a phenomenologically appealing range.

The remainder of this article is organised as follows: Section 2 begins with a summary of the model building constraints for the heterotic string with line bundles. We will then outline the modifications of the Green-Schwarz mechanism in the presence of five-branes and describe the M-theory origin of the new five-brane induced counter terms. Along the way we give an independent derivation, consistent with S-duality, of the correct normalisation of the Green-Schwarz terms, which we will need for concrete model building. This computation is relegated to Appendix A. The rest of section 2 discusses the issue of gauge-axion masses, threshold corrections and one-loop corrected Fayet-Iliopoulos terms and concludes with a brief discussion of the D-term potential for five-branes generated by line bundles. In section 3 we will introduce our novel realisation of the flipped $SU(5)$ including a brief discussion of Yukawa couplings and proton decay. Section 4 contains the results of our explicit model search on a Calabi-Yau manifold elliptically fibered over dP$_4$. We give the details of a vacuum with precisely the three-generation Standard Model chiral spectrum and no chiral exotics. Further such solutions which we have found are collected in Appendix B. Finally, in section 5, we describe how to directly arrive at the MSSM gauge group by a construction similar to that in section 4. Our model search again produces a couple of realistic solutions with just the MSSM chiral spectrum, as also displayed in Appendix B. Before concluding, we comment on the issue of gauge coupling unification in our scenario.    

{\bf Note added:} While we were preparing the results of this project for publication, we received the two very interesting eprints \cite{Tatar:2006dc} and \cite{Andreas:2006dm}, which also analyse aspects of heterotic vacua with abelian bundles.

\section{The $E_8\times E_8$ heterotic string with $U(N)$ bundles and five-branes}

We consider the heterotic $E_8\times E_8$ string compactified on a
Calabi-Yau manifold  $X$
endowed  with an additional vector bundle whose structure group is embedded
into the $E_8\times E_8$ ten-dimensional gauge group. In addition,
we also allow for the presence of heterotic respectively M-theory five-branes.

\subsection{Review of model building constraints}

To be more concrete, in this article we investigate  vector bundles of the 
following
form
\bea
\label{hbundle}
     W=W_1 \oplus    W_2,
\eea
where $W_1$ is embedded into the first $E_8$ and $W_2$ into the second.
For each of these bundles we take the Whitney sum 
\bea
\label{gbundle}
     W_i=V_{N_i} \oplus \bigoplus_{m_i=1}^{M_i}  L_{m_i} 
\eea
of the $U(N_i)$ bundle  $V_{N_i}$ and the complex line
bundles $L_{m_i}$. They are chosen such that 
\bea
\label{first0}
    c_1(W_i)=c_1(V_{N_i}) + \sum_{m_i=1}^{M_i} c_1(L_{m_i}) =0.
\eea
Consequently their structure group can be embedded into an $SU(N_i+M_i)$ 
subgroup of $E_8$. The observable gauge group, being  the commutant of
this structure group in $E_8^{(1)}$, is then given by
$H=E_{(9-N_i-M_i)}\times U(1)^{M_i}$. Depending on the line bundles it
can be further enhanced  or $U(1)$ factors can become massive
due to the Green-Schwarz mechanism.

Let us summarize some of the model building constraints to be satisfied \cite{Blumenhagen:2005ga}:

\begin{itemize}
\item{The vector bundles $W_i$ have to admit spinors, i.e.
\bea
\label{kthe}
       c_1(W_i)\in H^2(X,2\mbb{Z}),
\eea
which in  view of (\ref{first0}) is automatically guaranteed for the bundles we consider.} 

\item{The Bianchi identity for the three-form $H$
implies the tadpole cancellation condition
\bea
\label{TCC}
     0=  {1\over 4(2\pi)^2} 
 \left( {\rm tr}(\ov F_1^2)+{\rm tr}(\ov F_2^2)-{\rm tr}(\ov R^2)\right)
    -\sum_a N_a \ov\gamma_a,
\eea
to be satisfied in cohomology. Here $\ov F_i$ denotes the curvature two-form of the vector bundle $W_i$ and
$\ov \gamma_a$ represents the four-form Poincar\'e dual to the irreducible internal two-cycle $\Gamma_a$
wrapped by the $N_a$ M5-branes.
Note that the trace is over the fundamental representation of $E_8$ for the
gauge bundles and over the fundamental representation of $SO(1,9)$ 
for the curvature two-form. 

For the bundles (\ref{gbundle}) the resulting topological condition in cohomology can 
generally  be written as
\bea
\label{TCCgen}
           \sum_{i=1}^2 \left( {\rm ch}_2(V_{N_i}) + {1\over 2}\,
        \sum_{m_i=1}^{M_i}  c^2_1(L_{m_i}) \right)
           -\sum_a N_a \ov\gamma_a  =-c_2(T).
\eea
}

\item{At string tree level,  the connection of the vector bundle has to satisfy
the well-known zero-slope limit of the Hermitian Yang-Mills equations,
\bea
\label{HYM}
          F_{ab}=F_{\ov a\ov b}=0,\quad     g^{a\ov b}\, F_{a\ov b}=0.
\eea
The first equation implies that 
each term in (\ref{gbundle}) has to be a holomorphic vector bundle. 
The second equation can be satisfied precisely by holomorphic 
$\mu$-stable bundles which meet in addition the integrability condition
\bea
\label{DUY}        \int_{X}  J\wedge J \wedge c_1(V_{N_i}) = 0, \quad\quad
        \int_{X}  J\wedge J \wedge c_1(L_{m_i}) = 0,
\eea 
for all bundles individually. 
We will refer to the latter constraints in the sequel as the tree-level Donaldson-Uhlenbeck-Yau
(DUY) equation. 
As has been demonstrated in \cite{Blumenhagen:2005ga} for the case without any M5-branes,
the DUY condition arises in the effective
four-dimensional theory from a Fayet-Iliopoulus term for the
$U(1)$ gauge fields.  Moreover, it receives a one-loop correction,
which for the bundles we consider is of the form 
\bea
\label{DUYloop}
    &&\int_{X} J\wedge J \wedge c_1(L_{m_i})  - \\
    &&  \ell_s^4 \, g_s^2\,
       \int_{X} c_1(L_{m_i}) \wedge  \left( {\rm ch}_2(V_{N_i}) + {1\over 2}\,
        \sum_{n_i=1}^{M_i}  c^2_1(L_{n_i}) +
   {1\over 2}\, c_2(T)\right)=0 \nonumber
\eea
for all line bundles, where $g_s=e^{\phi_{10}}$ and $\ell_s=2\pi\sqrt{\alpha'}$. Of course the analogous condition  for $V_{N_i}$ instead of $L_{m_i}$ is to be considered as well.
 In section \ref{FI} we will compute how this equation
changes if M5-branes are included. In addition we have to require that the real part of the likewise loop-corrected gauge kinetic functions in the effective field theory are positive \cite{Blumenhagen:2005pm,Weigand:2005ng}. 
}
\end{itemize}

\subsection{Green-Schwarz mechanism with M5-branes}

The use  of non-trivial line bundles makes a careful analysis of anomalous $U(1)$ factors indispensible. It turns out that if we want to allow for the presence of five-branes to more easily find solutions of the tadpole condition (\ref{TCCgen}), the associated Green-Schwarz mechanism gets further modified. 
Let us explain  briefly  how the Green-Schwarz anomaly cancellation
mechanism works for the case that M5-branes are included. Along the way we will identify additional terms in the effective action which have to be present for a consistent coupling of the five-branes to the bulk action. 
Recall that for five-branes in the $SO(32)$ heterotic string,
there appears extra bifundamental chiral matter with respect
to the unitary gauge groups and
the symplectic gauge groups  supported on the 
five-branes. The resulting extra contributions
to the anomalies are cancelled by extra
Green-Schwarz terms from the 5-branes \cite{Blumenhagen:2005zg}. 

For the $E_8\times E_8$ heterotic string the story
must be different: In this case there does not exist chiral
matter arising from the M5-branes and in four dimensions
the M5-brane does not necessarily support a gauge field\footnote{If the M5-brane wraps a two-cycle in the Calabi-Yau
of genus $g$, then there exists a $U(1)^g$ gauge group in
four dimensions \cite{Lukas:1998hk}.}.

\begin{figure}
\begin{center}
\epsfxsize=5.9in 
\epsfbox{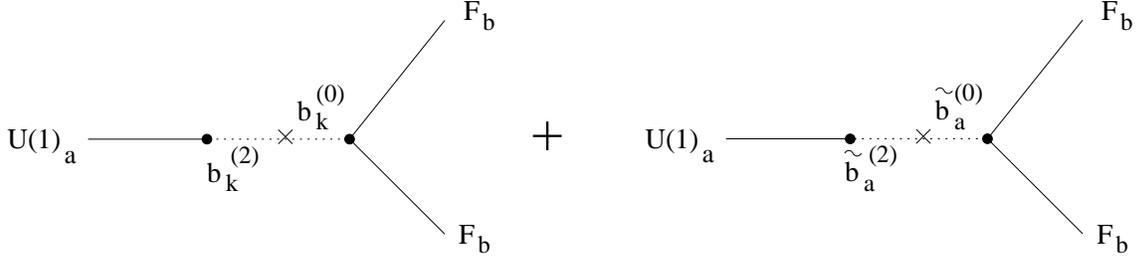}
\caption{Green-Schwarz counter term for the mixed gauge anomaly.
\label{figGS}}
\end{center}
\end{figure}

Unless stated otherwise we will be working in string frame and using the conventions of \cite{Polchinski:1998rr}. 
In ten dimensions the gauge and gravitational anomalies are cancelled by the counter term
\cite{Green:1984sg,Ibanez:1986xy}\footnote{There exists some confusion
in the literature about the correct normalization
of this term. In appendix A 
we will present an independent derivation from the S-dual
Type IIB orientifold point of view. Note in particular that this correct factor differs from the one used in \cite{Blumenhagen:2005ga,Blumenhagen:2005pm,Blumenhagen:2005zg,Weigand:2005ng,Honecker:2006dt} by $2$.}
\bea
     S_{GS}= {1\over 24\, (2\pi)^5\, \alpha'}\,  \int  B\wedge X_8,
\eea
where $B$ is the Kalb-Ramond two-form field and the eight-form
$X_8$ reads, as usual,
\bea   
  X_8={1\over 24} {\rm Tr} F^4 -{1\over 7200} \left( {\rm Tr} F^2\right)^2 
      -{1\over 240} \left( {\rm Tr} F^2\right) \left( {\rm tr} R^2\right)+
       {1\over 8}{\rm tr} R^4 +{1\over 32} \left( {\rm tr} R^2\right)^2 .
\eea
Explicitly taking care of the two $E_8$ factors by writing
$F=F_1+F_2$ one gets
\bea   
  X_8&=&{1\over 4} \left({\rm tr} F_1^2\right)^2+{1\over 4} \left({\rm tr} F_2^2\right)^2
   -{1\over 4} \left({\rm tr} F_1^2\right)\left({\rm tr} F_2^2\right)-
    {1\over 8} \left({\rm tr} F_1^2+{\rm tr} F_2^2\right)\left({\rm tr} R^2\right)+ \nonumber \\
 &&   {1\over 8}{\rm tr} R^4 +{1\over 32} \left( {\rm tr} R^2\right)^2 .
\eea
Using the tadpole cancellation condition~(\ref{TCC}), 
we dimensionally reduce this term to 
\bea
   S_{GS}&=& \sum_{i=1}^2 \biggl\{ 
   {1\over 8\, (2\pi)^3\, \alpha'}\, \int B\wedge \left({\rm tr}
     F_i^2\right)\, \left[ {1\over 4(2\pi)^2}\left({\rm tr} \ov F_i^2
         -{1\over 2} {\rm tr} \ov R^2\right) -{1\over 3} \sum_a N_a\, \ov
       \gamma_a \right]  \nonumber \\
      &+& {1\over 4\, (2\pi)^3\, \alpha'}\, \, \int B\wedge {\rm tr}( F_i\ov F_i ) 
\, \left[ {1\over 4(2\pi)^2}\left({\rm tr} \ov F_i^2
         -{1\over 2} {\rm tr} \ov R^2\right) -{1\over 3} \sum_a N_a\, \ov
       \gamma_a \right], \nonumber \\
     &+& \,{1\over 24\, (2\pi)^5\, \alpha'}\, \int B\wedge \left[{\rm tr}(
       F_i\ov F_i )\right]^2  \biggr\} \\
    &-&  {1\over 96\, (2\pi)^3\, \alpha'}\, \, \int B\wedge 
  \left[ {1\over 4(2\pi)^2}\left( {\rm tr} R^2\right)\left( {\rm tr} \ov
      R^2\right) -{2} \sum_a N_a\, \ov\gamma_a \right] \nonumber \\
   &-& \,{1\over 24\, (2\pi)^5\, \alpha'}\, \int B\wedge {\rm tr}( F_1\ov F_1
   ) \, {\rm tr}( F_2\ov F_2). \nonumber \\
\eea
In addition one has to take into account the kinetic term
\bea
\label{kinH}
         S_{kin}=-{1\over 4\kappa_{10}^2}\, \int e^{-2\phi_{10}}\,
       H\wedge \star_{10}\, H,
\eea
where $\kappa^2_{10}={1\over 2}(2\pi)^7\, (\alpha')^4$ and
the heterotic 3-form field  
strength  $H=dB^{(2)}-{\alpha'\over 4}(\omega_{Y}-\omega_{L})$ 
involves the 
gauge and gravitational Chern-Simons terms.
Following the steps detailed in \cite{Blumenhagen:2005ga} and computing the diagrams as shown in figure \ref{figGS}
results in the Green-Schwarz six-form
\bea
\label{ansix}
    A\sim A_{\rm pert} - {1\over 24 (2\pi)^4 \alpha'}\,
            \sum_a N_a \int_{\Gamma_a}  {\rm tr}( F_1\ov F_1)\, \biggl[
          {1\over 4}\left(  {\rm tr} F_1^2 + {\rm tr} F_2^2 - {\rm tr}
       R^2\right) + \nonumber \\
        {3\over 4}\left( {\rm tr} F_1^2 - {\rm tr} F_2^2 \right)
         \biggr] + (1\leftrightarrow 2). 
\eea
The term $A_{\rm pert}$  precisely  cancels the mixed abelian-nonabelian, mixed abelian-gravitational
and mixed abelian anomalies arising from the two $E_8$ factors \cite{Blumenhagen:2005ga}.
Since there does not exist any chiral matter from the M5-branes,
the only way to compensate the second term in (\ref{ansix}) is by additional
Green-Schwarz terms from the M5-branes. In fact, it is known that  on the six-dimensional
world-volume of an  M5-brane there lives a self-dual tensor 
field $\widetilde B_a$,
which by dimensional reduction gives rise to a two-form and a dual
scalar. These can generate additional Green-Schwarz counter terms.

More precisely, one can show that the following extra terms
\bea
\label{extragsa}
  S_{GS}^{(1)}={1\over 96 (2\pi)^3 \alpha'}\,
            \sum_a N_a \int_{\Gamma_a}  B \wedge
         \left(   {\rm tr} F_1^2 + {\rm tr} F_2^2 - {\rm tr}
       R^2\right) 
\eea
and 
\bea
\label{extragsb}
  S_{GS}^{(2)}={1\over 8   (2\pi)^3 \alpha'}\,
            \sum_a N_a \int_{\Gamma_a}  \widetilde B_a \wedge
         \left(   {\rm tr} F_1^2 - {\rm tr} F_2^2 \right) 
\eea
provide just the right counter terms to cancel the five-brane
dependent part in (\ref{ansix}). We would therefore like to argue that these terms have indeed to be present in the ten-dimensional effective action of the $E_8$ heterotic string for a consistent five-brane coupling. Even though the requirement of these terms by anomaly cancellation  is manifest only once we allow for background bundles with non-zero first Chern class, their presence cannot depend on the gauge instanton background, of course. In particular, they have an effect on the gauge kinetic function also of the field strength associated with the semi-simple part of the gauge group, as we will see in section 2.5.   
Note that the first term (\ref{extragsa}) involves the ten-dimensional
two-form $B$ and the second term (\ref{extragsb}) all the tensor
fields $\widetilde B_a$ living on the M5-branes.
The terms (\ref{extragsb}) can be viewed as being due to the cross terms in the kinetic action for the 
three-forms $\widetilde H_a$\footnote{The following normalization of  $\widetilde H_a$ arises after straightforward dimensional reduction of corresponding ten-dimensional terms to six dimensions. Due to the self-duality of $\widetilde H_a$ we should actually stick to the M-theory five-brane action \cite{Pasti:1997gx}, as will be done in section 2.3., but the resulting change of normalisation of the kinetic term is just a matter of field redefinitions.}
\bea
\label{kinHa}
     S_{kin} = -{1\over 2\,(2\pi)^3 (\alpha')^2} 
\int_{\Gamma_a} e^{-2\phi}\, \widetilde H_a\wedge \star_6\, \widetilde H_a
\eea
with 
\bea
         \widetilde H_a=d\widetilde B_a -{\alpha'\over 8} \left(
           \omega_{Y,1} -\omega_{Y,2} \right).
\eea
Both contributions to the effective action are also consistent with the analogous Green-Schwarz mechanism in six-dimensional compactifications, as analysed recently in \cite{Honecker:2006dt}.

\subsection{M-theory origin of new GS-terms}

The presence of these latter two counter terms (\ref{extragsa}) and (\ref{extragsb}) can also be
derived directly from Horava-Witten theory. The logic is very similar to that leading to the usual Green-Schwarz terms from heterotic M-theory, as first described in \cite{Lukas:1997fg,Lukas:1998ew}.
Here we will extend the analysis to the five-brane dependent terms. Our derivation was done independently from \cite{Carlevaro:2005bk}, where a similar analysis has been performed\footnote{Note that this reference does not use the resulting Green-Schwarz terms for cancellation of abelian anomalies and  does not consider the terms  (\ref{extragsb}) arising from the M5-brane action. Also, to the best of our knowledge, the connection between the new GS terms and the FI-D-terms in section 2.6 has not been explored previously.}.

To fix our conventions, the effective action of heterotic M-theory in the upstairs picture is given by \cite{Horava:1995qa, Horava:1996ma,Conrad:1997ww}
\bea
\label{SHW}
S&=& S_{kin} + S_{CS} + S_{curv} + S_{YM} + S_{M5}, \\\nonumber
S_{kin} &=&{1 \over 2\,\ov{\kappa}^2}  \, \int_{{\cal M}_u^{11}} \, R \, \Omega -{1 \over 2} G \wedge \star G,  \\\nonumber
S_{CS} &=& {1 \over 2\,\ov{\kappa}^2} \, \int_{{\cal M}_u^{11}} \frac{1}{6} C \wedge G \wedge G, \\\nonumber
S_{curv}&=&  {1 \over 48 (2 \pi)^3 \ov{\kappa}^2 T_5}  \int_{{\cal M}_u^{11}}  C \wedge \left(\frac{1}{8} {\rm tr} \,R^4 - \frac{1}{32} ({\rm tr} \,R^2)^2 \right),\\\nonumber
S_{YM}&=& -\sum_{i=1}^2\,{1 \over 2 \lambda^2} \int_{{\cal{M}}^{(10)}} {\rm tr}\,( F^{i} \wedge \star F^{i}) - {1 \over 2} {\rm tr} \,(R \wedge \star R),
\eea
where ${\cal M}_u^{11}={\cal{M}}^{10} \times S^{1}$. The part of M5-brane action $S_{M5}$ \cite{Pasti:1997gx} relevant for our purposes will be given at the end of this section.  The compact eleventh dimension takes values in the range $-\pi \rho <  x^{11} \leq  \pi \rho$ and the gauge fields are localized at $x^{11}= 0, \pi \rho$. It is understood that only the field components invariant under the orbifold action $x^{11} \mapsto - x^{11}$ are present in (\ref{SHW}) \cite{Horava:1995qa}. Eleven-dimensional indices will be denoted by $I,J,K,...$ and ten-dimensional ones by $A,B,C,...$. The ten-dimensional gauge couplings are related to $\ov{\kappa}$ via $\lambda^2 = (4 \pi) ( 2 \pi \ov{\kappa}^2)^{2/3}$ and the tension of the five-brane is given by $T_5= ({2 \pi \over \ov{\kappa}^4})^{1/3}$ \cite{Conrad:1997ww}.

In the presence of five-branes, the Bianchi identity for the field strength ${\mbox G=dC}$ gets modified as \cite{Lukas:1998hk}
\bea
(dG)_{11ABCD} = - {\ov{\kappa}^2 \over \lambda^2} \left( J_1 \delta( x^{11}) + J_2 \delta( x^{11}- \pi \rho) 
  + \frac{1}{2} J_5 (\delta( x^{11}- y)+\delta( x^{11}+ y)) \right)_{11ABCD}, \nonumber
\eea
taking into account the contribution from the five-brane at $x^{11}=y$ and its mirror brane at $x^{11}=-y$. The generalisation to the case of several five-branes is obvious.  The gauge and curvature sources at the orbifold fixed planes are given by $ {\mbox {}J_i= {\rm tr} \, F^{i} \wedge F^{i} - {1 \over 2} {\rm tr} \, R \wedge R = d\omega_i}$ for $i=1,2$, while the five-branes contribute ${\mbox {} J_5 = -4 (2\pi)^2 \delta(\Gamma)}$. 
Here $\delta(\Gamma)$ is the four-form Poincar\'e dual to the worldvolume of the five-brane in ${\cal M}^{(10)}$.\footnote{When we further compactify ${\cal M}^{(10)}={\cal M}^{(4)} \times CY_3$  we have the obvious decomposition  $ \delta(\Gamma)= \delta({\cal M}^{(4)}) \wedge \ov{\gamma}$ for a five-brane wrapping the two-cycle dual to the four-form  $\ov{\gamma}$ on $CY_3$.} In analogy with the Yang-Mills and Lorentz Chern-Simons forms we also introduce the ten-dimensional three-form $\omega_5$ satisfying ${\mbox{} J_5 = d\omega_5}$.

Being interested in the ten-dimensional theory after Kaluza-Klein reduction on $S^{1}$, we now focus on the situation where the eleventh dimension is much smaller than the ten-dimensional space. This is the limit in which the effective action of the ten-dimensional weakly coupled heterotic string arises \cite{Lukas:1997fg,Lukas:1998ew}.   In this regime ten-dimensional derivatives of gauge and curvature terms can be neglected as compared to field variations along $x^{11}$. Hence, one can give an approximate solution for $G$ and $C$ to the above Bianchi identity and the equations of motion $D^I \, G_{IJKL}=0$ by splitting the fields into their zero-mode and a background part as $C=C^{(0)} + C^{(1)}$ and $G=G^{(0)} + G^{(1)}$. Including also the five-brane sources, we get
\bea
\label{solB}
\begin{array}{cc}
C_{ABC}= C^{(1)}_{ABC}, &  C_{AB11}=B_{AB}, \nonumber\\ 
G_{ABCD}= G^{(1)}_{ABCD}, &    G_{ABC11}= (dB)_{ABC} + G^{(1)}_{ABC11},
\end{array}
\eea
\bea
C_{ABC}^{(1)}&=& - {\ov{\kappa}^2 \over 2 \lambda^2}\left( \omega_1 \epsilon(x^{11}) + \frac{1}{2} \omega_5 (\epsilon( x^{11}- y)+\epsilon( x^{11}+ y))- {x^{11} \over \pi \rho}(\omega_1 + \omega_2 + \omega_5) \right)_{ABC} \nonumber\\ 
G_{ABCD}^{(1)}&=& - {\ov{\kappa}^2 \over 2 \lambda^2}\left( J_1 \epsilon(x^{11}) + \frac{1}{2} J_5 (\epsilon( x^{11}- y)+\epsilon( x^{11}+ y))- {x^{11} \over \pi \rho}(J_1 + J_2 + J_5) \right)_{ABCD} \nonumber\\ 
G_{ABC11}^{(1)}&=& - {\ov{\kappa}^2 \over 2 \lambda^2 \, \pi \rho}(\omega_1 + \omega_2 + \omega_5)_{ABC}  . 
\eea
$\epsilon(x^{11})$ denotes the step function, i.e. $\epsilon(x^{11})= +1$ for $x^{11}$ positive and $-1$ otherwise. Note that the part of $G^{(1)}_{ABCD}$ linear in $x^{11}$ is cancelled precisely by the contribution from $G_{ABC11}^{(1)}$ when the Bianchi identity is evaluated. The equations of motion for the field strength $G$ are only satisfied up to terms proportional to $ \partial J_i$, which are assumed to be negligible in the limit we are considering  \cite{Lukas:1997fg,Lukas:1998ew}.

The ten-dimensional  weakly coupled heterotic string theory is recovered by compactification on $S^{1}$ according to the standard ansatz
\bea
ds_{11}^2 = e^{-2 \phi/3} \, g_{AB}^{(10)} \, dx^A \, dx^B + e ^{(4 \phi/3)} (dx^{11})^2.
\eea

In particular, the kinetic term for $G$ contains a part involving the combination $G_{11ABC} G^{11ABC}$. Inserting the solution (\ref{solB}), integrating over $S^{1}$ and focussing only on terms not involving $\omega_5$ due to the five-branes precisely yields the familiar kinetic term
\bea
\label{SkinH}
S_{kin}^H = -{1 \over 4 \kappa_{10}^2} \int_{{\cal M}^{(10)}} e^{-2 \phi} H \wedge \star H
\eea
for the ten-dimensional three-form field strength $H = dB - \frac{\alpha'}{4}(\omega_1 + \omega_2)$  
after setting
\bea
{1 \over  \kappa_{10}^2}= {2 \pi \rho \over \ov{\kappa}^2}, \quad \quad \quad \quad   \alpha' ={ 4\ov{\kappa}^2 \over 2 \lambda^2 \pi \rho} = { 2^{-1/3} \over \pi^2 \rho} 
({\ov{\kappa} \over 4 \pi})^{2/3}.
\eea    

We are now ready to investigate the origin of the complete Green-Schwarz counter terms including the contribution from the five-branes. They arise at order $({\ov{\kappa}^2 \over 2 \lambda^2})^2$ after inserting the above solution for $C$ and $G$ into the Chern-Simons terms $S_{CS}$  in (\ref{SHW}) as 
\bea
\label{CS}
S_{CS}|_{({\ov{\kappa}^2 \over 2 \lambda^2})^2} &=& {3 \over 12\,\ov{\kappa}^2}  \, \int_{{\cal M}^{(10)}} \int_{S^{1}} B \wedge G^{(1)} \wedge G^{(1)}\wedge dx^{11}\\
&=&{\pi \rho \over 4\ov{\kappa}^2} \,({\ov{\kappa}^2 \over 2 \lambda^2})^2 \int_{{\cal M}^{(10)}} B \wedge \left( {2 \over 3} (J_1^2 + J_2^2 - J_1 J_2 )- {1 \over 6} J_5 (J_1 + J_2)\right)\nonumber
\eea 
plus additional terms proportional to $\int B \wedge J_5^2$, which however vanish after performing the integral. To arrive at this expression we place the five-brane and its mirror symmetrically at $y=\pm{\pi \rho \over 2}$ between the two orbifold fixed-planes.
Additional contributions  from the higher curvature corrections $S_{curv}$ are 
\bea
\label{curv}
S_{curv}&=& {1 \over 48  (2 \pi)^3 \ov{\kappa}^2 T_5}  \int_{ {\cal M}_u^{11}} C \wedge \left (\frac{1}{8} {\rm tr} \,R^4 - \frac{1}{32} ({\rm tr} \,R^2)^2\right)\nonumber\\
&=& {1 \over 24 (2\pi)^5 \alpha'} \int _{{\cal M}^{(10)}} B \wedge \left(\frac{1}{8} {\rm tr} \,R^4 - \frac{1}{32} ({\rm tr} \,R^2)^2 \right).
\eea

The part ${2 \over 3} (J_1^2 + J_2^2 - J_1 J_2 )$  in (\ref{CS}) combines with (\ref{curv}) into the standard Green-Schwarz eight-form $X_8$ \cite{Lukas:1997fg,Lukas:1998ew}.

The additional counter terms (\ref{extragsa}) we are after now arise from $J_5 (J_1 + J_2) = -4 (2 \pi)^2 \delta(\Gamma) \wedge ({\rm tr} F_1^2 + {\rm tr} F_2^2 - {\rm tr} R^2)$.
In summary, (\ref{CS}) and (\ref{curv}) yield in the ten-dimensional limit
\bea
S_{GS}= c \int_{{\cal M}^{(10)}} B \wedge \left( X_8 + {(2 \pi)^2 \over 4}\, \delta(\Gamma)  \wedge ({\rm tr} F_1^2 + {\rm tr} F_2^2 - {\rm tr} R^2) \right) 
\eea
with 
\bea
c= {8 \over 3}{\pi \rho \over 4\ov{\kappa}^2} \,({\ov{\kappa}^2 \over 2 \lambda^2})^2 = {1 \over 24 (2 \pi)^5 \alpha'},
\eea
as postulated in (\ref{extragsa}).

The origin of the second five-brane dependent counter term (\ref{extragsb}) lies in the M5-brane action. With the normalisations of \cite{Pasti:1997gx} (see e.g. also \cite{Brandle:2001ts}), the part relevant for our analysis is given by
\bea
\label{M5}
S_{M5} =  -{T_5 \over 2} \sum_a N_a \int_{\Gamma_a \cup \Gamma_a'} \left({1\over 4}\widetilde{ F_a}\wedge \star \widetilde{F_a} + \tilde{C} + {1 \over 2} d{\widetilde{B}}_a \wedge C \right),
\eea
again summing over all branes and their mirrors. Here $\widetilde{F}_a=d\tilde{B}_a-C$ is the modified field strength of the self-dual tensor field $ \tilde{B}_a$ living on the five-brane and $\tilde{C}$ is the bulk six-form potential dual to $C$.
The contribution from (\ref{M5}) we are interested in is the topological coupling $d\tilde{B}_a \wedge C$. Following the general strategy we insert again the appropriate background solution for $C$ and place brane and mirror brane at $y= \pm {\pi \rho \over 2}$ respectively to find
\bea
S_{top}&=& - {T_5 \over 4}\sum_a N_a \left( \int_{\Gamma_a} \widetilde{B}_a \wedge dC^{(1)} + \int_{\Gamma'_a} \widetilde{B}_a \wedge dC^{(1)}\right)= \nonumber \\
&=& {T_5 \over 4}\, {\ov{\kappa}^2 \over 2 \lambda^2} \sum_a N_a \int_{\Gamma_a} \widetilde{B}_a \wedge ( {\rm tr} F_1^2 - {\rm tr} F_2^2).
\eea
It can be checked that, together with the kinetic term for $\widetilde{B}_a$, this coupling indeed yields precisely the required counter terms to cancel the contribution to the five-brane anomaly in the second line of (\ref{ansix}). In the standard ten-dimensional normalisation of the kinetic action for $\widetilde{B}_a$ which we used in (\ref{kinHa}) one eventually recovers the counter term  (\ref{extragsb}). Note that we are always free to change the normalization of $\widetilde{B}_a$; different normalizations simply correspond to different physical meanings of the modulus $\lambda_a$, to be introduced in section 2.5., which is associated to the position of the five-brane along $S^{1}$.

\subsection{Gauge-axion masses}

After this M-theoretic interlude we turn our attention to some relevant aspects of the four-dimensional effective action of our upcoming  construction of flipped 
$SU(5)\times U(1)_X$ and $SU(3)\times SU(2)\times U(1)_Y$
vacua. A central question we need to address is whether the abelian gauge factors really remain
massless after the Green-Schwarz mechanism cancels potential anomalies. For this purpose, we 
summarize in this section the various axion-gauge boson mass terms.

In the presence of five-branes, three kinds of axions come into the play: the universal axion, $b^{(0)}_0$,
the K\"ahler-axions, $b^{(0)}_k$, and finally the five-brane axions
$\widetilde b^{(0)}_a$.

They arise, together with their Hodge dual counterparts, from the reduction of the Kalb-Ramond form, its dual six-form and the self-dual two-forms on the five-branes  
\bea
    B^{(2)}&=& b^{(2)}_0+\ell_s^2\,\sum_{k=1}^{h_{11}}   b^{(0)}_k\, \omega_k ,\quad B^{(6)}=\ell_s^6\, b^{(0)}_0\,  {\rm vol}_6 + \ell_s^4\,
           \sum_{k=1}^{h_{11}}  b^{(2)}_k\, \widehat\omega_k,     \nonumber \\
\widetilde{B}_a&=& \widetilde b^{(2)}_a + \ell_s^2 \, \widetilde b^{(0)}_a \, \widehat\gamma_a.
\eea

Here we have introduced  a basis $\omega_k$, $(k= 1, \ldots, h_{11})$  of 
$H^2({X},{\mathbb Z})$ 
together with the Hodge dual four-forms $\widehat{\omega}_{k}$ normalised such that 
$\int_{X}  \omega_k  \wedge \widehat\omega_{k'} = \delta_{k, k'}$. Likewise $\widehat {\gamma}_a$ is the Hodge dual of ${\ov\gamma}_a$.  
Furthermore ${\rm vol}_6$ represents the volume form of the internal Calabi-Yau manifold normalised such 
that $\int_{X} {\rm vol}_6 = 1$.

The final piece of notation we will need occurs in  expressing relevant traces  as
\bea
\label{traces}
     {\rm tr}_{E_8^{(i)}}(F_i \ov F_i)&=&\sum_{m_i,n_i=1}^{M_i} 
\kappa_{m_i,n_i}\, f_{m_i}\, \ov f_{n_i} \\\nonumber
     {\rm tr}_{E_8^{(i)}}(F_i^2)&=& 2\,  {\rm tr}_{E_{9-N_i-M_i}}(F_i^2) + 
\sum_{m_i,n_i=1}^{M_i} \eta_{m_i,n_i}\, f_{m_i}\, f_{n_i}, 
\eea
where the $f_{m_i}$ and $\ov f_{m_i}$ denote the observable and internal $U(1)$ gauge fields. Note that the non-abelian part of the traces carries different numerical pre-factors for bundles other than the one used in this article (see e.g. \cite{Honecker:2006dt}).
It is then a straightforward task to extract the mass terms for the various axions  (or equivalently their dual two-forms) from the heterotic action \cite{Blumenhagen:2005ga}:
The mass terms for the universal axion  reads
\bea
\label{massa}
  M_{0,m_i}&=& {1\over 4(2\pi)^2\alpha'}  \int_{{\cal M}^{(4)}} b^{(2)}_0\wedge
     f_{m_i} \Biggl[\sum_{n_i=1}^{M_i} \kappa_{m_i,n_i} 
   \int_X c_1(L_{n_i})\wedge 
     \Biggl( {\rm ch}_2(V_{N_{i}})   + \nonumber \\ 
     &&   {1\over 2}\, \sum_{k_{i}=1}^{M_{i}}    c^2_1(L_{k_{i}}) +
   {1\over 2}\, c_2(T)   
  -{1\over 4}\sum_a N_a \ov\gamma_a \Biggr) \Biggr].
\eea
For the K\"ahler axions the kinetic term for $H_3$ induces the
mass terms
\bea
\label{massb}
  M_{k,m_i}&=& {1\over 2(2\pi)^2\alpha'}  \int_{{\cal M}^{(4)}} b^{(2)}_k\wedge
     f_{m_i} \Biggl[\sum_{n_i=1}^{M_i} \kappa_{m_i,n_i} 
   \int_X c_1(L_{n_i})\wedge \widehat \omega_k   \Biggr]
\eea
and the 5-brane Green-Schwarz term yields the mass term
\bea
\label{massc}
  M_{a,m_i}&=& \pm {1\over 4\, (2\pi)^2\alpha'}  \int_{{\cal M}^{(4)}} 
    \widetilde b^{(2)}_a\wedge
     f_{m_i} \Biggl[\sum_{n_i=1}^{M_i} \kappa_{m_i,n_i} 
   \int_X c_1(L_{n_i})\wedge  \ov\gamma_a   \Biggr]
\eea
for the 5-brane axions. The plus sign holds for $E_8^{(1)}$ and
the minus sign for $E_8^{(2)}$. 

A combination of abelian gauge fields or of axions, respectively, remains
massless, if it lies in the kernel of this whole  axion-gauge boson mass matrix $(M_0, M_k, M_a)_{m_i}$.

\subsection{Gauge couplings}

In this section we extract the holomorphic gauge kinetic functions  for
the non-abelian and abelian gauge groups \cite{Derendinger:1985cv,Choi:1985bz,Ibanez:1986xy,Nilles:1997vk,Stieberger:1998yi}. 
The tree-level gauge kinetic function for the non-abelian factor is still simply $f=S$ in terms of
the complexified dilaton 
\bea
    S&=&{1\over 2\pi}\left[ e^{-2\phi_{10}} {{\rm Vol}({X}) \over
             \ell_s^6 } + {i}\, b^{(0)}_0 \right]. 
\eea
Let us furthermore define 
\bea
     {1\over 4(2\pi)^2} {\rm tr} \ov F_i^2 = \sum_{k=1}^{h_{11}}  ({\rm tr} \ov F_i^2)_k\,\,
        \widehat \omega_k,     
   \quad \quad  \quad \quad \ov \gamma_a = \sum_{k=1}^{h_{11}}  ( \ov \gamma_a)_k\,\,
        \widehat \omega_k.     
\eea
Then the one-loop corrected gauge kinetic function for the non-abelian
gauge fields in the large radius regime can be written as
\bea
\label{gaugekin1}
     f_{1,2}=S+{1\over 8}\sum_{k=1}^{h_{11}} T_k\, {\left({\rm tr} \ov F_{1,2}^2
         -{1\over 2} {\rm tr} \ov R^2- \sum_a N_a \ov\gamma_a\right)_k}
            \pm {1\over 2} \sum_a N_a\, \Lambda_a.
\eea
The complex scalars appearing above are the bosonic part
of the ${\cal N}=1$ superfields and given by     
\bea
         T_k&=&{1\over 2\pi} \left[- {1\over \ell^2_s}\int_X J\wedge \widehat\omega_k
           + i b^{(0)}_k \right], \\
         \Lambda_a&=&{1\over 2\pi} \left[ -{\lambda_a}\, 
             {{\rm Vol}(\Gamma_a)\over
            \ell_s^2} +i \widetilde b^{(0)}_a \right]. 
\eea           
The $\lambda_a$ denote the scalars which together with the self-dual
two-forms $\widetilde B_a$ combine into  tensor multiplets on the
six-dimensional world-volume of the five-branes. 
In the strong coupling Horava-Witten model these scalars are nothing
else than the position of the respective five-branes along the 
eleventh direction.

\begin{figure}
\begin{center}
\epsfxsize=3.6in 
\epsfbox{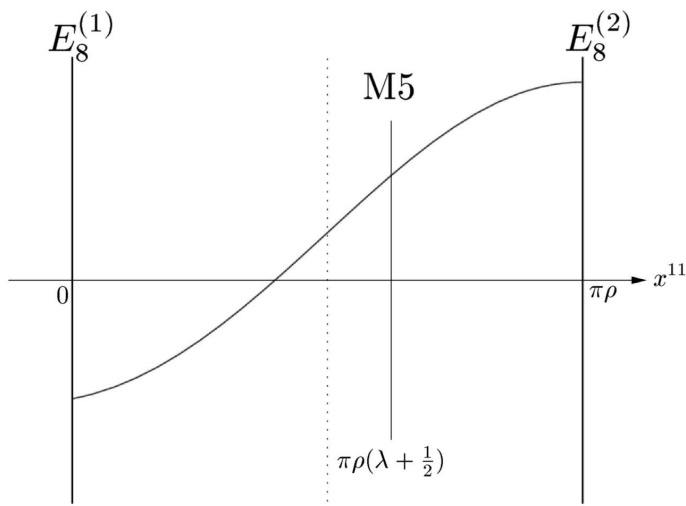} 
\caption{M5-brane potential in Horava-Witten theory on the Quintic induced by abelian gauge flux on $E_8^{(1)}$. 
\label{fighorava}}
\end{center}
\end{figure}

In fact, what one can easily read off from the dimensionally reduced heterotic action are the couplings between the various axions and the gauge fields and thus the imaginary part of (\ref{gaugekin1}). The relevant contributions stem again partly from the Green-Schwarz terms and the cross-terms of the kinetic action for the underlying two-form fields (see again \cite{Blumenhagen:2005ga} for the details in our context). It is of course the power of four-dimensional ${\cal N}=1$ supergravity which allows us to simply reconstruct the full expression (\ref{gaugekin1}) since the gauge kinetic functions are holomorphic.  
Consequently, for the  real part, i.e. for the gauge couplings, one gets at linear order in $\lambda_a$
\bea
     {4\pi\over  g^2_{1,2}} &=&{e^{-2\phi_{10}} \over 3\ell_s^6}\int_X J\wedge J\wedge J
              - {1\over  \ell_s^2}
              \int_X J\wedge {1\over 4(2\pi)^2} \left( 
                {\rm tr} \ov F^2_{1,2} -{1\over 2} {\rm tr} \ov
         R^2\right)\nonumber \\
           &&   + {1\over  \ell_s^2} \sum_a N_a\, \left( {1\over 4} \mp 
                 \lambda_a \right)\,
                 \int_{\Gamma_a} J. 
\eea
The first term is the tree-level gauge coupling and receives one-loop
theshold corrections depending both on the K\"ahler moduli of 
the Calabi-Yau and the five-brane moduli $\lambda_a$ (see also \cite{Carlevaro:2005bk}).
If we set all five-brane moduli to zero, then we nevertheless get a five-brane contribution
of $1/4$ to the one-loop gauge couplings in both the first and the second $E_8$.
From the Horava-Witten point of view this means that for   $\lambda_a=0$,
the five-brane is placed exactly in the middle between the two end of the world nine-branes
and $\lambda_a$ is measured with respect to this symmetric configuration (see figure
\ref{fighorava}). We will give further evidence for this interpretation momentarily. 

The next-to-leading order M-theory computation carried out in \cite{Derendinger:2000gy, Moore:2000fs}
provides an  ${\cal O}(\lambda^2)$ correction to the real part of the dilaton superfield
\bea
\label{S}
    S&=&{1\over 2\pi}\left[ e^{-2\phi_{10}} {{\rm Vol}({X}) \over
             \ell_s^6} +\sum_a N_a\, {\lambda_a^2\over 2\ell_s^2} 
    \int_{\Gamma_a} J + {i}\, b^{(0)}_0 \right]. 
\eea
This correction was derived in \cite{Moore:2000fs} essentially by requiring that the kinetic terms for the self-dual two-form on the M5-brane can indeed be correctly incorporated into an appropriate K\"ahler potential. Using this result and holomorphy of the gauge kinetic function leads to the gauge couplings
\bea
\label{gaugekinfun1}
     {4\pi\over  g^2_{1,2}} &=&{1\over 3 \ell_s^6\,  g_s^2}\int_X J\wedge J\wedge J
              - {1\over  \ell_s^2}
              \int_X J\wedge \Biggl( {\rm ch}_2(V_{N_{1,2}}) + {1\over 2}\,
        \sum_{n_{1,2}=1}^{M_{1,2}}  c^2_1(L_{n_{1,2}}) + \nonumber \\
   && {1\over 2}\, c_2(T)\Biggr) 
     + {1\over  \ell_s^2} \sum_a N_a\, \left( {1\over 2} \mp 
                 \lambda_a \right)^2\,
                 \int_{\Gamma_a} J. 
\eea
Note that the one-loop threshold corrections for the non-abelian gauge groups 
are universal inside each $E_8$ wall. 
For $\lambda_a = -{ 1 \over 2}$, the contribution of the five-brane to the threshold corrections from  $E_8^{(1)}$ is precisely that of a small instanton inside $E_8^{(1)}$. This unambiguously identifies $\lambda_a$ as the relative position of the five-brane measured with respect to the middle of the interval between the orbifold planes, as suggested already. We point out once more that different normalisations of the counter terms (\ref{extragsb}) would have resulted in a corresponding redefinition of $\lambda_a$.  
Moreover, as expected, 
if one places the five-brane
inside the $E_8^{(2)}$ wall, its gauge threshold corrections to the gauge couplings 
from $E_8^{(1)}$ vanish and vice versa. 

For the abelian gauge groups things are slightly different.
The resulting in general non-diagonal abelian gauge couplings are given by
\bea
\label{gaugekinfun2}
   4\pi {\rm Re}( f_{m_i,n_i}) &=&{\eta_{m_i,n_i}  \over 12 \ell_s^6 g_s^2}\, \int_X J\wedge J\wedge J
              - {\eta_{m_i,n_i}\over 4 \ell_s^2}
              \int_X J\wedge \Biggl( {\rm ch}_2(V_{N_{i}}) + {1\over 2}\,
        \sum_{k_i=1}^{M_{i}}  c^2_1(L_{k_i}) + \nonumber \\
   && {1\over 2}\, c_2(T) \Biggr)  -{1\over 12 \ell_s^2}
              \int_X J\wedge \left( \sum_{p_i,q_i=1}^{M_i} \kappa_{m_i,p_i}\kappa_{n_i,q_i} \,
             c_1(L_{p_i})\, c_1(L_{q_i}) \right) \\
    &&  + {\eta_{m_i,n_i}\over 4 \ell_s^2} \sum_a N_a\, \left( {1\over 2} \mp 
                 \lambda_a \right)^2\,
                 \int_{\Gamma_a} J  \nonumber
\eea
for both $U(1)$ factors from the same $E_8$ factor and by
\bea
\label{gaugekinfun3}
  4\pi {\rm Re}(f_{m_1,n_2} ) ={4\pi\, f_{m_1,n_2} } =  {1\over 24 \ell_s^2}
              \int_X J\wedge \left(  \sum_{p_1=1}^{M_1}  \sum_{q_2=1}^{M_2} \kappa_{m_1,p_1}\kappa_{n_2,q_2} \,
             c_1(L_{p_1})\, c_1(L_{q_2}) \right)
\eea
for one $U(1)$ from the first and one $U(1)$ from the second $E_8$ (cf (\ref{traces}) for notation).
Apparently, only for trivial line bundles, i.e. Wilson lines, do the extra threshold corrections
vanish.

\subsection{Fayet-Iliopoulos terms}
\label{FI}
Whenever we are dealing  with  anomalous $U(1)$ gauge factors, there
are potential Fayet-Iliopoulos (FI) terms generated (e.g.\cite{Dine:1987xk}).
With the help of the standard supersymmetric field theory formula
\bea
\label{FIterms}
  D_m\, {\xi_m\over g_m^2}=D_m\,   {\partial {\cal K}\over \partial V_m}
\biggr\vert_{V=0},
\eea
the FI parameters $\xi_m$  can be computed from the K\"ahler potential
${\cal K}$. The K\"ahler potential in turn is of course determined by requiring that it reproduces the various kinetic terms in the four-dimensional action in the Einstein frame. The kinetic terms of the M5-brane action do not only induce the quadratic correction (\ref{S}) in the definition of the superfield $S$, but also dictate a modification of the standard contribution  $-{\rm ln}(S+ S^*)$ to ${\cal K}$  by a term quadratic in $\Lambda_a +\Lambda_a^*$ \cite{Derendinger:2000gy, Moore:2000fs}. Furthermore, the various axionic couplings of the abelian gauge fields induce additional modifications of ${\cal K}$ such that these terms are recovered in the superfield formalism in the same way as the kinetic terms. In supergravity language, the rationale behind this is that the ${\cal N}=1$ supergravity gets gauged by the axionic shift symmetry; by a standard procedure the K\"ahler potential has to be rendered  invariant under this gauging. As a result, the correct K\"ahler potential takes the form
\bea
{\cal K} &=&-{M^2_{pl}\over 8\pi}    
       \ln\Biggl[S+S^*-\sum_m Q^m_0\, V_m +
   \sum_a {N_a\over 2} {(\Lambda_a +\Lambda_a^* -\sum_m Q^m_a\, V_m)^2\over
                        (\ov\gamma_a)_k  (T_k+T_k^* - \sum_a Q^m_k\, V_m) }\Biggr]  \nonumber \\
    &&-{M^2_{pl}\over 8\pi} \ln\Biggl[-\sum_{i,j,k=1}^{h_{11}}
{d_{ijk}\over 6} 
\biggl(  T_i+T_i^*-\sum_m Q^m_i\, V_m\biggr)  \biggl(  T_j+T_j^*-\sum_m
Q^m_j\, V_m\biggr) \nonumber \\
&& \phantom{aaaaaaaaaaa} \biggl(  T_k+T_k^*-\sum_m Q^m_k\, V_m\biggr) \Biggr].
\eea
Here ${M_{pl}^2\over 8\pi}=\kappa_{10}^{-2}\, {\rm Vol}(X)$ and $V_m$ denotes the vector superfields. Note that for $V_m=0$ this is essentially the result derived in \cite{Derendinger:2000gy, Moore:2000fs}.
The charges $Q^m_k$
can be identified as the couplings in the mass terms (\ref{massa}),(\ref{massb}),(\ref{massc})
using the definition
\bea
                 S_{mass}=\sum_{m=1}^M  \sum_{k=0}^{h_{11}} 
            {Q^m_k\over 2\pi\alpha'}
                \int_{\IR_{1,3}} f_m\wedge b^{(2)}_k+
        \sum_{m=1}^M  \sum_{a} 
            {Q^m_a\over 2\pi\alpha'}
                \int_{\IR_{1,3}} f_m\wedge \widetilde b^{(2)}_a. 
\eea

We finally obtain for the Fayet-Iliopoulos term 
\bea
\label{fayet}
    {\xi_{m_i}\over g_{m_i}^2}& =  & - \frac{1}{ 8 \ell_s^6} \sum_{n_i=1}^{M_i} \kappa_{m_i,n_i}
     \Biggl[
        \int_{X} J\wedge J \wedge 
         {\ov f_{n_i} \over 2\pi}  - {e^{2 \phi_{10}} }\, {\ell_s^4 }
       \int_{X} {\ov f_{n_i}\over 2\pi} \wedge
          {1\over 4(2\pi)^2} {\left({\rm tr} \ov F_i^2
         -{1\over 2} {\rm tr} \ov R^2\right)}  \nonumber \\
    && \phantom{aaaaaaa} + {e^{2 \phi_{10}} }\,{\ell_s^4 } \sum_a N_a 
       \left( {1\over 2} \mp \lambda_a \right)^2\,
                 \int_{\Gamma_a} {\ov f_{n_i} \over 2\pi} \Biggr].
\eea
Thus we learn that a flux through the two-cycle
$\Gamma_a$ 
of a five-brane on the wall $E_8^{(i)}$ generates a one-loop D-term potential for the
five-brane modulus $\lambda_a$. From (\ref{fayet}) it seems at first sight that this D-term repels the five-brane from the wall
and vanishes only if the five-brane lies on top of the other 
wall.  We will come back to this issue in section 2.7.

The FI-term (\ref{fayet}) generalizes the one-loop corrected DUY equation in
the presence of M5-branes to
\bea
\label{DUYlooph5}
    &&\int_{X} J\wedge J \wedge c_1(L_{m_i})  - 
      \ell_s^4 \, g_s^2\,
       \int_{X} c_1(L_{m_i}) \wedge  \biggl( {\rm ch}_2(V_{N_i}) + {1\over 2}\,
        \sum_{n_i=1}^{M_i}  c^2_1(L_{n_i}) + \nonumber\\
  && {1\over 2}\, c_2(T)\biggr) +  \ell_s^4 \, g_s^2\ 
    \sum_a N_a  \left( {1\over 2} \mp 
                 \lambda_a \right)^2\,
                 \int_{\Gamma_a} c_1(L_{m_i}) =0.
\eea
In general, these conditions provide constraints fixing combinations of the
K\"ahler moduli, the dilaton and five-brane moduli.

Let us pause a moment to comment on the physical interpretation of the five-brane contribution to the above D-term. Arising at one loop in the weakly coupled heterotic string, it is expected to be due to appropriate amplitudes from membranes after unfolding the wrapped eleventh dimension in the strongly-coupled Horava-Witten regime. In fact, as derived in \cite{Moore:2000fs}, there are non-perturbative contributions to the F-term superpotential from open membranes stretching between one of the orbifold fixed planes and the M5-brane provided that the worldvolume of the membranes is precisely of the form $I \times \Gamma_a$. Here $I$ simply denotes the interval along the eleventh dimension between the orbifold plane and five-brane and $\Gamma_a$ is as always the internal cycle wrapped by the M5-brane. We see that, apparently, such configurations also contribute to the D-term potential if the membrane can couple to some abelian background gauge flux on the orbifold plane. As is manifest in (\ref{DUYlooph5}), this can only happen if the five-brane wraps a two-cycle which, pulled back to the end of the world, carries non-vanishing gauge flux. In particular, this interpretation explains why the five-brane is sensitive to the presence of the gauge flux along $\Gamma_a$ even though it may be placed at an arbitrary position along the eleventh dimension: The presence of the gauge flux is communicated by the exchange of appropriate open membranes.

\subsection{D-term potential for M5-branes}

Let us discuss in a simple example what the effect of the FI terms
is for the M5-brane modulus $\lambda$.
It is instructive to consider the Quintic Calabi-Yau manifold, which
has only one K\"ahler modulus, and we assume that we have chosen
a vector bundle $V\oplus L^{-1}$ embedded into the first $E_8$ 
wall without any matter charged under the $U(1)$.
Then the D-term potential arising from the FI-term of the $U(1)$ is simply
\bea
         V_D={1\over 2} g^2\, \left( {\xi\over g^2} \right)^2,
\eea
where $g$ denotes the gauge coupling of the $U(1)$.
For the Quintic one has $c_2(T)=10\eta^2$ and $J=\ell_s^2 r\, \eta$
with $r>0$ in terms of the single $(1,1)$-form $\eta$. Moreover, we write ${\rm ch_2}(V)=-v\eta^2+{1\over 2}l^2\eta^2$ and 
${\rm ch}_2(L)={1\over 2}l^2\eta^2$ and introduce one five-brane
wrapping the class $\gamma$. The tadpole cancellation condition then
reads
\bea
         -v+l^2 -\gamma^2=-10.
\eea
The relevant D-term potential takes the form
\bea
   V_D\simeq { \left( {r^2\over g_s^2}- (\gamma^2-5) + \left({1\over
           2}-\lambda\right)^2 \, \gamma^2  \right)^2 \over
           \left( {r^2\over g_s^2}- 3 (\gamma^2-5) + 3 \left({1\over
           2}-\lambda\right)^2 \, \gamma^2 -{\kappa_{1,1}^2\over \eta_{1,1}}
       \, l^2  \right)}.
\eea
For fixed string coupling $g_s=0.5$, radius $r=2$ and parameters $\gamma=l=2$, 
${\kappa_{1,1}^2/\eta_{1,1}}=1/10$, this potential for
the five-brane modulus $\lambda$ has the characteristic shape shown in figure \ref{fighorava}.
Naively from the FI-term one might have expected that the five-brane is repelled by the $E_8$ walls
carrying a non-trivial line bundle. However, the contribution
of the $g^2$ term multiplying
the FI-term in the scalar potential changes this picture and leads to an attractive potential between
the five-brane and the $E_8$ wall carrying the bundle.
This is actually well in agreement with the interpretation of the D-term potential as being due to open membranes stretching  between the orbifold fixed plane and the M5-brane: Their contribution is of course minimized precisely if the interval along which they wrap between the end of the world and the five-brane is vanishing.

\section{Flipped $SU(5)\times U(1)_X$ }

We are finally in a position to discuss the central part of this article, the application of the formalism introduced so far to the construction
of GUT like heterotic string compactifications.
As pointed out in the introduction, what has been discussed in detail so far in the literature is the following standard realisation of GUT gauge groups. On the Calabi-Yau one chooses
an $SU(4)$ or $SU(5)$ bundle embedded into one of the two $E_8$ factors. 
The resulting observable gauge groups are $SO(10)$ or $SU(5)$,
respectively. In addition one gets chiral matter transforming in the
${\bf (16)}$ or  ${\bf (10)+(\ov 5)}$ representation of the gauge group.
However, in these scenarios there does not appear a Higgs field that can
break the GUT group down to the Standard Model. This is achieved
by breaking   $SO(10)$ or $SU(5)$ via non-trivial discrete Wilson lines, which in general 
can only exist if the fundamental group of the Calabi-Yau is non-trivial. 
Such Calabi-Yau's can be constructed by taking free discrete quotients
of a Calabi-Yau with vanishing fundamental group. 
The electroweak Higgs can appear from the ${\bf (10)}$ or the ${\bf (5)+(\ov 5)}$
representations. 
Without doubt, from the physical point of view, this is a very simple and compelling picture and
recently models whose particle spectrum is quite close to the 
Standard Model have been constructed \cite{Donagi:2000zs,Braun:2005nv,Bouchard:2005ag}. 

As mentioned above, in this construction the breaking of the gauge symmetry down
to the Standard Model is achieved via discrete Wilson lines, in more mathematical
terms by flat abelian bundles. In this section we would like
to investigate whether one can use also non-flat line bundles
to obtain phenomenologically interesting GUT models. 
We will see that we will almost directly be led to considering flipped $SU(5)$ models.

\subsection{$SU(4)\times U(1)$ bundles}

Instead of first using an $SU(4)$ bundle and then breaking

$SO(10)$ down to flipped  $SU(5)$ by a discrete Wilson line,
we consider a bundle with structure group $SU(4)\times U(1)$
on a Calabi-Yau $X$ with $\pi_1(X)=0$. 
Such types of construction have been considered in \cite{Distler:1987ee}
before and further details of this particular one can be found in \cite{Blumenhagen:2005ga} (see also the recent articles \cite{Tatar:2006dc,Andreas:2006dm}). 

The starting point is a bundle
\bea    \label{BundleU4a}
    W=V\oplus L^{-1}, \quad {\rm with}\ c_1(V)=c_1(L), \, \, \,  \mbox{rank}(V)=4, 
\eea
with structure group $SU(4)\times U(1)$.
This bundle $W$ can now
be embedded into an $SU(5)$ subgroup of $E_8$ so that the commutant
is  $SU(5)\times U(1)_1$.  
We embed the $U(1)$ bundle such that the charge matrix is
\bea
    {\cal Q}_1=(1,1,1,1,-4),
\eea
in the notation of \cite{Blumenhagen:2005ga}.
In fact, consider the breaking of the original structure group 
$SU(5) \rightarrow U(4)\times U(1)$ and the corresponding decomposition of 
$(\5,{\bf 10}) \rightarrow (\4,{\bf 10})_{-1} + (\1, {\bf 10})_4$ to read off 
the unique charge assignments of $V$ and $L$.
Using the decomposition of the adjoint representation of $E_8$ 
\bea
\label{breaking}
{\bf 248} 
 \stackrel{SU(4) \times SU(5) \times U(1)}{\longrightarrow}
\left\{\begin{array}{c}
({\bf 15},\1)_0 \\
(\1, \1)_0 + (\1,{\bf 10})_4 + (\1, \ov{\bf 10})_{-4} + (\1, {\bf 24})_0 \\
(\4,\1)_{-5} + (\4, \ov{\5})_3 + (\4,{\bf 10})_{-1} \\
(\ov{\4},\1)_{5} + (\ov{\4}, \5)_{-3} + (\ov{\4},\ov{\bf 10})_{1} \\
(\6,\5)_2 + (\6,\ov{\5})_{-2}
\end{array}\right\} ,
\eea
we list the resulting massless
spectrum 
in Table 1.
\begin{table}[htb]
\renewcommand{\arraystretch}{1.5}
\begin{center}
\begin{tabular}{|c||c|c|}
\hline
\hline
$SU(5)\times U(1)_1$ & bundle & SM part. \\
\hline \hline
${\bf 10}_{-1}$ & $V$ & $(q_L,d^c_R,\nu^c_R)+[H_{10}+\ov H_{10}]$ \\
${\bf 10}_{4}$ & $L^{-1}$ &$-$ \\
\hline
$\overline \5_{3}$ & $V\otimes L^{-1}$ & $(u^c_R,l_L)$ \\
 $\overline \5_{-2}$ & $\bigwedge^2 V$ & $[(h_3, h_2)+(\ov h_3, \ov h_2)]$ \\
\hline
 ${\bf 1}_{-5}$ & $V\otimes L$ & $e^c_R$ \\
\hline
\end{tabular}
\caption{\small Massless spectrum of $H=SU(5)\times U(1)_1$ models.}
\label{signsb}
\end{center}
\end{table}

\noindent
The massless fields precisely carry, up tp a common factor, the 
$U(1)_X$ charges as appearing in the flipped $SU(5)$ \cite{Barr:1981qv,Derendinger:1983aj}, $Q_X=-\frac{1}{2} \, Q_1$.\footnote{Note that the normalisation of $Q_X$, as chosen here, differs from the one in \cite{Barr:1981qv} by a factor of  $\frac{1}{2}$.}
Recall that this model differs from the conventional Georgi-Glashow GUT scenario in that the $U(1)_Y$ is not entirely contained in the $SU(5)$, but arises as the specific linear combination
\bea
\frac{1}{2}Q_Y = -\frac{1}{5}Q_Z + \frac{2}{5}Q_X,
\eea
where $Z$ is the generator of $SU(5)$ commmuting with the generators of the Standard Model  $SU(3) \times SU(2)$. In the normalisation of \cite{Barr:1981qv} $Z$ is given by 
$Z={\rm diag}(-1/3,-1/3,-1/3,1/2,1/2)$.
The way how the MSSM matter organizes into flipped $SU(5)$ multiplets is related to the Georgi-Glashow scenario by the flip
\bea
d^c_R \leftrightarrow u^c_R, \quad\quad e^c_R \leftrightarrow \nu^c_R.
\eea
Most importantly, the $({\bf 10})_{-1}$ contains the right-handed neutrino as a particle uncharged under the MSSM $SU(3) \times SU(2)\times U(1)_Y$, and giving it a VEV  can therefore serve as the Higgs effect breaking the GUT group down to the Standard Model one. It is this peculiarity of flipped $SU(5)$ which at first sight allows us to work on manifolds without Wilson lines.
However, if we only consider the  bundle (\ref{BundleU4a}) inside the first $E_8$ with $c_1(L)\ne 0$, one K\"ahler/dilaton modulus receives  a mass from the DUY constraint and therefore
also one axion in combination with the $U(1)_X$ gauge boson.
Therefore, after GUT Higgsing by $H_{10}$ the resulting $U(1)_Y$ would
also be massive. 
This seems to bring us back into the old situation that we are forced to consider  manifolds with non-vanishing
fundamental group to allow for non-trivial flat bundles\footnote{
For $\pi_1(X)=0$, a line bundle with $c_1(L)=0$ is always trivial
and the observable gauge group gets enhanced to $SO(10)$.}. 

Alternatively, here we propose to embed a second line bundle
into the other $E_8$, such that a linear combination of the two observable
$U(1)$'s remains massless.
Concretely, we embed the line bundle $L$ (or more precisely $L \oplus L^{-1}$) also in the second $E_8$, where it leads 
to the breaking
$E_8\to E_7 \times U(1)_2$ and the decomposition
\bea
\label{breaking}
{\bf 248} 
 \stackrel{E_7 \times U(1)}{\longrightarrow}
\left\{\begin{array}{c}
({\bf 133})_0 +(\1)_0 + ({\bf 56})_1 +({\bf 56})_{-1} + ({\bf 1})_{2}
 + ({\bf 1})_{-2} \\
\end{array}\right\} .
\eea
The resulting massless
spectrum is displayed in Table \ref{signsd}.
\begin{table}[htb]
\renewcommand{\arraystretch}{1.5}
\begin{center}
\begin{tabular}{|c||c|c|}
\hline
\hline
$E_7\times U(1)_2$  & bundle \\
\hline \hline
${\bf 56}_{1}$ & $L^{-1}$ \\
${\bf 1}_{2}$ & $L^{-2}$  \\
 \hline
\end{tabular}
\caption{\small Massless spectrum of $H=E_7\times U(1)_2$ models.}
\label{signsd}
\end{center}
\end{table}
\noindent
The tadpole cancellation condition for this model reads
\bea
\label{tadpolecond}
      {\rm ch}_2(V)+3\, {\rm ch}_2(L)-\sum_a N_a\ov \gamma_a=-c_2(T).
\eea
Considering the mass terms in (\ref{massa}) and (\ref{massb}) and computing
\bea
      \kappa_{1,1}=-10, \quad\quad \kappa_{2,2}=-4,
\eea
one realizes that the linear combination
\bea
           U(1)_X=-{1\over 2}\left( U(1)_1 -{5\over 2}\, U(1)_2 \right)
\eea
remains massless if the following conditions are satisfied
\bea
\label{masslesscond}
         \int_X  c_1(L)\wedge c_2(V) =0, \quad\quad  \int_{\Gamma_a} c_1(L)=0 \quad {\rm for \, \,  all\, \,  M5\, \, branes}.
\eea
 The resulting chiral massless spectrum simplifies considerably
and is  given in Table 3.
\begin{table}[htb]
\renewcommand{\arraystretch}{1.5}
\begin{center}
\begin{tabular}{|c||c|c|}
\hline
\hline
$SU(5)\times U(1)_X \times E_7$ & bundle & SM part. \\
\hline \hline
$ ({\bf 10},{\bf 1})_{1\over 2}$ & $\chi(V)=g$ & $(q_L,d^c_R,\nu^c_R)+[H_{10}+\ov H_{10}]$ \\
$( {\bf 10},{\bf 1})_{-2}$ & $\chi(L^{-1})=0$ &$-$ \\
\hline
$(\overline \5,{\bf 1})_{-{3\over 2}}$ & $\chi(V\otimes L^{-1})=g$ & $(u^c_R,l_L)$ \\
$(\overline \5,{\bf 1})_{1}$ & $\chi(\bigwedge^2 V)=0$ & $[(h_3, h_2)+(\ov h_3, \ov h_2)]$ \\
\hline
 $({\bf 1},{\bf 1})_{{5\over 2}}$ & $\chi(V\otimes L)+\chi(L^{-2})=g$ & $e^c_R$ \\
\hline
$({\bf 1},{\bf 56})_{{5\over 4}}$ & $\chi(L^{-1})=0$ & $-$ \\
\hline
\end{tabular}
\caption{\small Massless spectrum of $H=SU(5)\times U(1)_X$ models 
      with $g={1\over 2}\int_X c_3(V)$.}
\label{signsb}
\end{center}
\end{table}

\noindent
Therefore, one gets precisely $g$ generations of flipped $SU(5)$ matter. 
Note that in general the right-handed electrons receive contributions from both the 
first and the second $E_8$. From a phenomenological point of view, we want to circumvent these latter in order to avoid non-MSSM like selection rules for their Yukawa couplings. They are absent if additionally one requires
\bea
\label{rhelectrons}
          \int_X c^3_1(L)=0.
\eea
If the tadpole condition is satisfied, this implies, together with (\ref{masslesscond}), that also $\int_X  c_1(L)\wedge c_2(T) =0$.
With these extra conditions, the generalized DUY condition for the bundle $L$
simplifies considerably,
\bea
\label{modelDUY}
       \int_X J\wedge J\wedge c_1(L)=0, 
\eea
and contains only the tree-level part. Therefore, it freezes only one of
the $h_{11}$ K\"ahler moduli. By contrast, the threshold corrections to the gauge kinetic functions will be non-vanishing. For consistency of the low-energy effective theory we need to ensure that the DUY can actually be solved in a regime inside the K\"ahler cone where the real part of the threshold corrected gauge kinetic functions is positive, at least for the unbroken gauge symmetries. Apart form the $SU(5)$ and the hidden $E_7$ symmetry, we will therefore have to check this condition for the gauge kinetic function of the generator of $U(1)_X$, which is given by
\bea
\label{gaugekinU(1)}
f_{X,X}= {1 \over 4} \left( f_{1,1} + \left({5 \over2}\right)^2\,  f_{2,2} + 5 \, f_{1,2} \right)
\eea 
in terms of the the corresponding objects for $U(1)_1$ and $U(1)_2$.

\subsection{Yukawa couplings and proton decay}

This string theory realization of flipped $SU(5)\times U(1)_X$ features many, though not all, 
of the characteristic features of the field theory GUT model. For their details we refer to \cite{Barr:1981qv,Derendinger:1983aj,Antoniadis:1987dx}. 
In particular, the GUT breaking via a non-vanishing vacuum expectation value of
the singlet component in $H_{10}+\ov H_{10}$ leads to a natural
solution of the doublet-triplet splitting problem via a missing partner mechanism
in the superpotential  coupling
\bea
                   {\bf 10}^H_{{1\over 2}}\,  {\bf 10}^H_{{1\over 2}}\, {\bf 5}_{-1} .
\eea 
Therefore,  problematic dimension-five operators mediating proton decay can be suppressed. 
Moreover, the gauge invariant Yukawa couplings
\bea
     {\bf 10}^i_{{1\over 2}}\,  {\bf 10}^j_{{1\over 2}}\, {\bf 5}_{-1}, \quad
     {\bf 10}^i_{{1\over 2}}\,  {\bf \ov 5}^j_{-{3\over 2}}\, {\bf \ov 5}_{1},\quad
     {\bf \ov 5}^i_{-{3\over 2}}\, {\bf 1}^j_{{5\over 2}}\, {\bf 5}_{-1}, \
 \eea
lead to Dirac mass-terms for the $d$, $(u,\nu)$ and $e$ quarks and leptons after electroweak
symmetry breaking. If there exist additional gauge singlets $\Phi$, then couplings of the form
${\bf 10}^i_{{1\over 2}}\, {\bf \ov {10}}^H_{-{1\over 2}}\, \Phi$ can give rise
to Majorana type neutrino masses and therefore to a see-saw mechanism.

Since the electroweak Higgs carries different quantum numbers than the lepton doublet, 
the dangerous dimension-four proton decay operators 
\bea
 {\bf l}\,{\bf l}\,{\bf e} \quad  \in \quad   {\bf \ov 5}^i_{-{3\over 2}}\, {\bf 1}^j_{{5\over 2}}\, {\bf \ov 5}^k_{-{3 \over 2}}, \quad \quad \quad \quad
{\bf q}\,{\bf d}\,{\bf l}, \quad {\bf u}\,{\bf  d}\,{\bf d} \quad \in \quad  {\bf 10}^i_{{1\over 2}}\,  {\bf 10}^j_{{1\over 2}}\, {\bf \ov 5}^k_{-{3 \over 2}} \quad 
\eea
are not gauge invariant and thus absent. A detailed discussion of this peculiar property of heterotic constructions with line bundles has recently been given in \cite{Tatar:2006dc} in the context of Georgi-Glashow $SU(5)$. Furthermore, as shown in \cite{Dorsner:2004jj}, flipped $SU(5)$ differs from the Georgi-Glashow model in that also the dimension-six proton decay operators, emerging after integrating out the off-diagonal gauge bosons in the $({\bf 3}, {\bf 2})$, can be completely eliminated. Additional details and more references can also be found in \cite{Nath:2006ut}.   

The issue of gauge coupling unification for this model is absolutely identical
to the discussion to be presented in section \ref{gaugeunif}. Due to the embedding of
$U(1)_X$ into both $E_8$ walls, the tree level  gauge couplings satisfy
the relation $\alpha_3=\alpha_2={8\over 3}\alpha_Y$ instead of the usual factor $5/3$.
Therefore, with  the minimal supersymmetric matter content the tree level gauge couplings 
do not unify at $M_{GUT}=2\cdot 10^{16}$GeV. What rescues us are the one-loop threshold corrections to be further
discussed in section \ref{gaugeunif}.

\section{Three-generation models}
\label{example}
It is now time to construct explicit examples of our framework with three chiral generations. This will be done choosing as the background manifold  an elliptically fibered Calabi-Yau threefold, where we can exploit the spectral cover construction \cite{Friedman:1997ih,Friedman:1997yq} to endow it with stable holomorphic vector bundles. Let us first recapitulate briefly the most important aspects of such a construction (see also the account in \cite{Donagi:1999gc} for further details). The reader familiar with this technique may wish to jump directly to section 4.3.

\subsection{Elliptically fibered Calabi-Yau manifolds}
An elliptically fibered Calabi-Yau three-fold $X$ consists of a complex two-surface $B$ as base space, together with an analytic map
\bea
\pi: X \rightarrow B,
\eea
where fibers over each point $b$ in the base
\bea
E_b=\pi^{-1}(b)
\eea
are elliptic curves, which can be described by the homogeneous Weierstrass equation
\bea
zy^2 = 4x^3-g_2xz^2-g_3z^3.
\eea
Additionally, we require $X$ to admit a global section $\sigma: B \rightarrow X$, assigning to every point in the base $b\in B$ the zero element $\sigma(b)=p \in E_b$ on the fiber.
We denote by $\mathcal{L}$ the conormal bundle to the section $\sigma(B)$. One can regard $g_2$ and $g_3$ as sections of $\mathcal{L}^4$ and $\mathcal{L}^6$, respectively, where $\mathcal{L}^i$ is the $i$-fold tensor product of $\mathcal{L}$ with itself.

The Calabi-Yau condition requires, besides the fact that the base space and the fibers have to be complex manifolds themselves, that the first Chern class of the tangent bundle $T$ must vanish,
\bea
c_1(T)=0.
\eea
This implies $\mathcal{L}=K_B^{-1}$, where $K_B$ is the canonical bundle of the base space. It follows that $K_B^{-4}$ and $K_B^{-6}$ must have sections $g_2$ and $g_3$, respectively. The surfaces compatible with this condition are found to be del Pezzo, Hirzebruch, Enriques und blow-ups of Hirzebruch surfaces.

It has been shown that on such spaces the Chern classes of the tangent bundle of the total space are given by the Chern classes of the base space. Especially, we need later the second Chern class of the tangent bundle:
\bea
\label{tangentbundle}
c_2(T)=12\sigma c_1(B)+11c_1(B)^2+c_2(B),
\eea
where $\sigma$ is the Poincar\'e dual two-form to the global section.

\subsection{Spectral cover construction}
A $U(n)$ or $SU(n)$ vector bundle over an elliptically fibered Calabi-Yau space $X$ can be obtained via the so-called spectral cover construction. The starting point is the observation that a vector bundle on an elliptic curve splits into a direct sum of line bundles
\bea
{\cal V}=\mathcal{N}_1\oplus\ldots\oplus\mathcal{N}_n.
\eea
Since finally, the vector bundle should be supersymmetric, it must admit a field strength  satifying the hermitian Yang-Mills equations, which is equivalent to the requirement that the associated bundle is holomorphic  $\mu$-semi-stable. Therefore, the line bundles $\mathcal{N}_i$ must be of degree zero. On an elliptic curve this means that there is a unique point $Q_i$ such that there is a meromorphic section of $\mathcal{N}_i$ with a pole at $Q_i$ and with a zero at the origin $p$. Thus, the vector bundle on the curve is determined by an unordered $n$-tuple of points on the curve.

In intuitive terms, on an elliptically fibered space a vector bundle is then determined by a set of $n$ points, varying over the base, i.e. an $n$-fold cover of $B$ with $\pi_C: C \rightarrow B$. This is called the spectral cover, determining the vector bundle on the fibers. One has additionally to specify one more line bundle $\mathcal{N}$ such that $\pi_{C\ast}\mathcal{N}$ is a vector bundle on $B$ with the direct sum of $n$ lines $\mathcal{N}|_{Q_i}$ as fibers. The whole vector bundle on $X$ is given using the Poincar\'e bundle $\mathcal{P}$, which is a line bundle over the fiber product $X\times_B C$,
\bea
{\cal V}=\pi_{1\star}(\pi_2^{\ast}\mathcal{N}\otimes\mathcal{P}).
\eea
Here $\pi_1$ and $\pi_2$ are the projections on the first and second factor of the fiber product, respectively.

Since for $U(n)$ bundles constructed in this way, the semi-stability is an intricate issue, we consider from now on only $SU(n)$ bundles and twist them finally with $U(1)$ bundles to obtain $U(n)$ bundles, which are $\mu$-semi-stable again
. For an $SU(n)$-bundle we have the additional condition that the first Chern class must vanish. Friedman, Morgan and Witten \cite{Friedman:1997yq} have found a formula for $c_1({\cal V})$,

\bea
c_1({\cal V})=\pi_{C\ast}(c_1(\mathcal{N})+\tfrac{1}{2}c_1(C)-\tfrac{1}{2}\pi^\ast c_1({\cal V})),
\eea
where $c_1(C)$ and $c_1(B)$ are the first Chern classes of the tangent bundles of $C$ and $B$, respectively. Since $c_1({\cal V})$ must vanish, we have
\bea
c_1(\mathcal{N})=-\tfrac{1}{2}c_1(C)+\tfrac{1}{2}\pi_C^{\ast} c_1(B) + \gamma
\eea
in terms of the  cohomology class $\gamma$ satisfying
\bea
\pi_{C\ast}\gamma=0.
\eea
In general, $\gamma$ can be writen as
\bea
\gamma=\lambda(n\sigma-\pi_C^{\ast}\eta+n\pi_C^{\ast}c_1(B)),
\eea
where $\lambda\in\mathbb{Q}$. Putting everything together, we have
\bea
\label{linebundle}
c_1(\mathcal{N})=n(\tfrac{1}{2}+\lambda)\sigma + (\tfrac{1}{2}-\lambda)\pi_C^{\ast}\eta + (\tfrac{1}{2}+n\lambda)\pi_C^\ast c_1(B).
\eea
Since $c_1(\mathcal{N})$ must be an integer class, not every value of $\lambda$ is allowed. One often uses the conditions:
\begin{align}
n\text{\quad odd}:\quad &\lambda\in\mathbb{Z}+\tfrac{1}{2},&\nonumber\\
n\text{\quad even}:\quad &\lambda\in\mathbb{Z},&\quad  \eta+c_1(B)=0 \mod 2,\\
&\lambda\in\mathbb{Z}+\tfrac{1}{2},&\quad  c_1(B)=0 \mod 2.\nonumber
\end{align}
Note that these are only sufficient conditions. In our example, we also consider the more general case that $\lambda$ can be of higher fraction.

The $SU(n)$ bundles constructed so far are only $\mu$-semi-stable. It has been shown \cite{Friedman:1997ih} that the spectral cover must be irreducible in order to obtain a $\mu$-stable one\footnote{In
  fact, the proof of stability assumes that the K\"ahler parameter of the
  fiber lies in a certain range near the boundary of the 
  K\"ahler cone, i.e $J=\epsilon \sigma +J_B$ with
sufficiently small $\epsilon$. Since the value of $\epsilon$ is not
known, in all models involving the spectral cover constructions it is
  therefore a subtle issue if the region of stability overlaps with the
  perturbative regime, which is needed to have control over non-perturbative
  effects. In all examples which will be relevant for us, the constraints will leave us enough freedom to go to regions of the K\"ahler cone where $\epsilon$ is much smaller than $J_B$.}, which imposes two more conditions to the curve $\eta$ \cite{Donagi:2004ia}:
\begin{itemize}
\item The linear system $|\eta|$ has to be base point free.
\item The class $\eta - nc_1(B)$ has to be effective.
\end{itemize}

A $\mu$-stable $U(n)$ bundle can be obtained by twisting an $SU(n)$ bundle with an additional line bundle $\mathcal{Q}$ on $X$ with
\bea
c_1(\mathcal{Q})=q\sigma+c_1(\zeta),
\eea
where $c_1(\zeta) \in H^2(B,\mathbb{Z})$. The resulting $U(n)$ bundle
\bea
V={\cal V}\otimes\mathcal{Q}
\eea
is $\mu$-stable if the original $SU(n)$ bundle is. Using $\ch({\cal V}\otimes\mathcal{Q})=\ch({\cal V})\ch(\mathcal{Q})$, we can compute the resulting Chern characters of $V$ from the ones of ${\cal V}$ \cite{Curio:1998vu,Andreas:1998ei} and $\mathcal{Q}$:
\bea
\ch_1(V)&=&nq\sigma+nc_1(\zeta),\\
\ch_2(V)&=&\left[-\eta+\frac{nq}{2}(2c_1(\zeta)-qc_1(B))\right] \sigma + a_F,\\
\ch_3(V)&=&\lambda\eta(\eta-nc_1(B))-\eta c_1(\zeta)+q\left(\frac{n}{2}c_1(\zeta)^2-\omega\right)+\\
	&&qc_1(B)\left(\eta-\frac{nq}{2}c_1(\zeta)+\frac{nq^2}{6}c_1(B)\right).\nonumber
\eea
Here
\bea
a_F&=&\frac{n}{2}c_1(\zeta)^2-\omega,\\
\omega&=&-\frac{1}{24}  c_1(B)^2 (n^3-n) + \frac{1}{2} \left( \lambda^2 - \frac{1}{4}\right) n\eta\ (\eta-nc_1(B)),
\eea
where $\ch_3(V)$ has already been integrated over the fiber.
The Chern classes read
\bea
c_1(V)&=&nq\sigma+nc_1(\zeta),\\
c_2(V)&=&\left[\eta+nq(n-1)\left(c_1(\zeta)-\frac{q}{2}c_1(B)\right)\right]\sigma+\frac{n^2}{2}c_1(\zeta)^2-a_F,\\
c_3(V)&=&\frac{nq^2}{6}(n^2-3n+2)\left(qc_1(B)^2-3c_1(\zeta)c_1(B)\right),\\
&&+\frac{nq}{2}(n^2-2n+2)c_1(\zeta)^2+(2q-nq-2n\lambda)\eta c_1(B)\nonumber\\
&&+(n-2)\eta c_1(\zeta)+2\lambda\eta^2-nqa_F-2q\omega.\nonumber
\eea
To summarize, a $U(n)$ bundle is completely specified by the rational number $\lambda$, the integer $q$ and the classes $\eta$ and $c_1(\zeta)$.

\subsection{An example on $\dP_4$}
The del Pezzo surfaces $\dP_r$, $r=0,1,\ldots,8$ are obtained from the complex projective space $\mathbb{P}^2$ by blowing up $r$ points. Hence $H^2(\dP_r)$ is generated by $r+1$ elements: $l$, which is the class inherited from the projective space, and $E_1,\ldots,E_r$, generated by the blown-up points. The intersection numbers are
\bea
l\cdot l &=& \int_{\dP_r}l \wedge l = 1,\nonumber\\
E_i\cdot E_j &=& \int_{\dP_r}E_i \wedge E_j = -\delta_{ij},\\
E_i\cdot l &=& \int_{\dP_r}E_i \wedge l = 0,\nonumber
\eea
and the Chern classes read
\bea
c_1(\dP_r)=3l-\sum_{m=1}^r E_m, \quad c_2(\dP_r)=3+r.
\eea
The effective classes on del Pezzo surfaces form a cone, called Mori cone. The Mori cone is linearly generated by a finite set of effective classes if $r\leq8$. The linear system of a curve $|\eta|$ is base point free if $\eta\cdot E\geq 0$ for every curve $E$ with $E^2=-1$ and $E\cdot c_1(B)=1$, which are precisely the generators of the Mori cone.

We turn now concretely to $\dP_4$ as the base space for an elliptically fibered Calabi-Yau space $X$. The second Chern class of the tangent bundle is given by \eqref{tangentbundle},
\bea
c_2(T)=[36l-12\sum_{i=1}^4E_i]\, \sigma+62F,
\eea
where $c_1(\dP_4)$ is expanded in the cohomological basis and $F$ is the class of the fiber. The Mori cone is generated by the 10 effective classes $E_i$, $l-E_i-E_j$, $i,j=1,\ldots,4$, $i\neq j$.

We have finally introduced all the relevant technology for constructing interesting solutions. We have found a couple of three-generation flipped $SU(5)$ vacua satisfying all the required constraints. They are listed in Table 5 of Appendix B. We choose the following example to demonstrate their properties.
The $U(4)$ bundle is given by the data
\bea
\lambda&=&\frac{1}{4},\quad q=0,\nonumber\\
\eta&=&14l-2E_1-6E_2-6E_3-2E_4,\\
c_1(\zeta)&=&-l+E_2+E_3+E_4.\nonumber
\eea
Note that the first Chern class of the line bundle $\mathcal{N}$ in the spectral cover constructon \eqref{linebundle} is an integer class, as required:
\bea
c_1(\mathcal{N})=3\sigma + \pi_C^\ast \left (8l - 2E_1 - 3E_2 -3E_3 - 2E_4 \right).
\eea
It is easy to see that $|\eta|$ is base point free, since its intersection with the generators of the Mori cone is always positive. One can also easily show that $\eta$ is effective as well as $\eta-4c_1(\dP_4)=2l+2E_1-2E_2-2E_3+2E_4$. Thus, this bundle is $\mu$-stable. The resulting Chern classes are
\bea
c_1(V)&=&-4l+4E_2+4E_3+4E_4,\\
c_2(V)&=&[14l-2E_1-6E_2-6E_3-2E_4]\, \sigma-29F.
\eea
In our setup, the first Chern class of the line bundle must be equal to the first Chern class of the vector bundle (see (\ref{BundleU4a})), thus 
\bea
c_1(L)=-4l+4E_2+4E_3+4E_4.
\eea

In our example model, we also include M5-branes.  Their combined associated cohomology class is
\bea
[W] =  27F + (22l - 10E_1 - 6E_2 - 6E_3 - 10E_4) \, \sigma.
\eea
To make physical sense, $[W]$ must be Poincar\'e dual to the homology class of a curve $\Gamma$ in $X$, and must be therefore effective. $[W]$ is effective if its part on the fiber is greater than or equal to zero and its part on the base is effective in $B$. Therefore, we rewrite $[W]$ in terms of generators of the Mori cone,
\bea
\label{fivebranedecomp}
[W]&=&\sum_aN_a\overline{\gamma}_a=27F+[12E_1+6(l-E_1-E_2)\\
&&+6(l-E_1-E_3)+10(l-E_1-E_4)]\, \sigma\nonumber.
\eea
The generators of the Mori cone, being irreducible as effective classes, represent the classes dual to the irreducible curves $\gamma_a$ around which we wrap $N_a$ five-branes. In general,
this decomposition is in general not unique. However, we also have to satisfy the constraint ${ \mbox{}\int_{\Gamma_a}c_1(L)=0}$ for a  massless $U(1)_X$, and \eqref{fivebranedecomp} is the only remaining decomposition compatible with this requirement.
The tadpole cancellation condition \eqref{TCCgen} for this setup, written in terms of Chern classes, takes the form
\bea
-c_2(V)+2c_1^2(L)-[W]=-c_2(T)
\eea
and is indeed satisfied.
It is a simple calculation to show that the conditions to keep the $U(1)_X$ in the flipped $SU(5)$ model massless hold
\bea
\int_X c_1(L) \wedge c_2(V)=0, \quad \int_{\Gamma_a}c_1(L)=\int_X c_1(L)\wedge \overline{\gamma}_a=0.
\eea
Since the Chern class of the line bundle has no part in the fiber, the integral over its third power trivially vanishes,
\bea
\int_X c_1^3(L)=0,
\eea
and thus a contribution to the right-handed electrons from the second $E_8$ factor is prevented. 
The number of generations in our example is given by
\bea
\chi(V)=\frac{1}{2}\int_X c_3(V) = 3
\eea
since  $\int_Xc_1(V)\wedge c_2(V)=\int_Xc_1(L)\wedge c_2(V)=0$.

Expanding the K\"ahler class in the cohomological basis,
\bea
J=l_s^2(r_\sigma\sigma+r_0l+\sum_{m=1}^4 r_mE_i),
\eea
the DUY-equation \eqref{modelDUY} 
\bea
\label{myduy}
\int_X J\wedge J\wedge c_1(L)=-8l_s^4r_\sigma(r_0+r_2+r_3+r_4) = 0
\eea
fixes one K\"ahler modulus. There exist solutions inside the K\"ahler cone. Take as an example
\bea
0 < r_\sigma < 2\rho,\quad r_0=3\rho,\quad r_m=-\rho,\quad m=1,\ldots,4.
\eea
With this choice, equation \eqref{myduy} holds and the K\"ahler class lies inside the K\"ahler cone for every $\rho\in\mathbb{R}^+$.

The universal gauge coupling for the non-abelian visible gauge group (\ref{gaugekinfun1}) can be computed as\footnote{Note that in the following equations, $\lambda_5$ is the five-brane modulus and not the parameter belonging to the bundle data.} 
\bea
\frac{4\pi}{g_1^2}=\frac{1}{3g_s^2}\left(5r_\sigma^3-15r_\sigma^2\rho+15r_\sigma\rho^2\right)-24r_\sigma-4\rho-(\tfrac{1}{2}-\lambda_5)^2(7r_\sigma-34\rho),
\eea
which is positive for a suitable choice of parameters. The abelian gauge couplings are given by (\ref{gaugekinfun2},\ref{gaugekinfun3})
\bea
4\pi\preal(f_{i,i})&=&\frac{\eta_{i,i}}{4}\left(\frac{1}{3g_s^2}(5r_\sigma^3-15r_\sigma^2\rho+15r_\sigma\rho^2)\right.\\
&&\left.-24r_\sigma-4\rho-(\tfrac{1}{2}-\lambda_5)^2(7r_\sigma-34\rho)\right)+\frac{320}{3}r_\sigma,\nonumber\\
4\pi\preal(f_{1,2})&=&-\frac{160}{3}r_\sigma
\eea
with $\eta_{1,1}=40$ and $\eta_{2,2}=4$.
The resulting gauge coupling (\ref{gaugekinU(1)}) for the $U(1)_X$ is then positive again:
\bea
4\pi\preal f_{X,X} &=& \frac{65}{16}\left(\frac{1}{3g_s^2}(5r_\sigma^3-15r_\sigma^2\rho+15r_\sigma\rho^2)-24r_\sigma-4\rho\right.\\
&&\left.-(\tfrac{1}{2}-\lambda_5)^2(7r_\sigma-34\rho)\right)+\frac{380}{3}r_\sigma.\nonumber
\eea

To summarize, this example with three chiral generations satisfies the tadpole condition (\ref{tadpolecond}) as well as the constraints (\ref{masslesscond}) guaranteeing a massless $U(1)_X$. We have no non-MSSM like selection rules for the Yukawa couplings of the right-handed electrons since there are indeed no contributions from the second $E_8$ (\ref{rhelectrons}). Furthermore, the K\"ahler moduli can be chosen such that the DUY equation (\ref{modelDUY}) holds and the gauge couplings are positive.

In Appendix B, we list all three-generation models we have found on dP$_4$ by a computer search which likewise satisfy all these conditions. We have also found three-generation examples for a scenario directly giving rise to the Standard Model gauge symmetry, to be discussed in the next section.

\section{Just the $SU(3)\times SU(2)\times U(1)_Y$  gauge symmetry}

Since the Standard Model contains an abelian gauge symmetry, one can try
to get the Standard Model gauge symmetry directly from embedding
abelian bundles into $E_8\times E_8$. The philosophy is actually very similar to the one presented in section~3.

\subsection{$SU(5)\times U(1)$ bundles}

The direct breaking of $E_8$ to the Standard Model group is indeed possible by choosing a bundle with structure
group $SU(5)\times U(1)$.
Now we start with a bundle
\bea    \label{BundleU4}
    W=V\oplus L^{-1}, \quad {\rm with}\ c_1(V)=c_1(L), \, \, \,  \mbox{rank}(V)=5, 
\eea
which has structure group $SU(5)\times U(1)$.
This bundle $W$ can
be embedded into an $SU(6)$ subgroup of $E_8$ such that the commutant
is  $SU(3)\times SU(2)\times U(1)_1$.  
We embed the $U(1)$ bundle such that
\bea
    {\cal Q}_1=(1,1,1,1,1,-5).
\eea
The decomposition of the adjoint representation of $E_8$ reads
\bea
\label{breaking}
{\bf 248} 
 \stackrel{SU(5) \times SU(3)\times SU(2) \times U(1)_1}{\longrightarrow}
\nonumber
\left\{\begin{array}{c}
({\bf 24}; \1, \1)_0 + (\1; \1, \1)_0 + (\1;{\bf 8}, \1)_0 +(\1;\1,{\bf 3})_0 \\
({\bf 5}; {\bf 3},{\bf 2})_1 + (\1; {\bf 3},{\bf 2})_{-5} + c.c.\\
({\bf 10}; {\bf {\ov 3}},{\bf 1})_2 + ({\bf 5}; {\bf \ov 3},{\bf 1})_{-4} + c.c.\\
({\bf \ov {10}}; {\bf 1},{\bf 2})_3 + ({\bf 5}; {\bf 1},{\bf 1})_{6} + c.c.
\end{array}\right\}
\eea
and apparently contains states with just the Standard Model quantum numbers.
Here again the $U(1)_1$ by itself cannot remain massless so that we will
perform the same construction as for the flipped $SU(5)$ model.
We embed the line bundle also in the second $E_8$ and realize that here
the linear combination
\bea
           U(1)_Y={1\over 3}\left( U(1)_1 + 3\, U(1)_2 \right)
\eea
remains massless if again the conditions 
\bea
         \int_X  c_1(L)\wedge c_2(V) =0, \quad\quad  \int_{\Gamma_a} c_1(L)=0 
\eea
are satisfied.
The resulting chiral massless spectrum takes the simple form  given in Table 4.
\begin{table}[htb]
\renewcommand{\arraystretch}{1.5}
\begin{center}
\begin{tabular}{|c||c|c|}
\hline
\hline
$SU(3)\times SU(2)\times U(1)_Y \times E_7$ & bundle & SM part. \\
\hline \hline
$ ({\bf 3},{\bf 2}, {\bf 1})_{1\over 3}$ & $\chi(V)=g$ & $q_L$ \\
$ ({\bf 3},{\bf 2}, {\bf 1})_{-{5\over 3}}$ & $\chi(L^{-1})=0$ & $-$ \\
\hline
$ ({\bf \ov 3},{\bf 1}, {\bf 1})_{2\over 3}$ & $\chi(\bigwedge^2 V)=g$ & $d^c_R$ \\
$ ({\bf \ov 3},{\bf 1}, {\bf 1})_{-{4\over 3}}$ & $\chi(V\otimes L^{-1})=g$ & $u^c_R$ \\
\hline
$ ({\bf 1},{\bf 2}, {\bf 1})_{-1}$ & $\chi(\bigwedge^2 V\otimes L^{-1})=g$ & $l_L$ \\
$ ({\bf 1},{\bf 1}, {\bf 1})_{2}$ & $\chi(V\otimes L)+\chi(L^{-2}) =g$ & $e^c_R$ \\
$({\bf 1},{\bf 1},{\bf 56})_{1}$ & $\chi(L^{-1})=0$ & $-$ \\
\hline
\end{tabular}
\caption{\small Massless spectrum of $H=SU(3)\times SU(2)\times U(1)_Y$ models 
      with $g={1\over 2}\int_X c_3(V)$.
}
\label{signsb}
\end{center}
\end{table}

\noindent
Therefore, one gets precisely $g$ generations of Standard Model matter without a right-handed neutrino. 
The right-handed electrons have contributions from both the 
first and the second $E_8$. The latter are again absent if additionally one requires
\bea
          \int_X c^3_1(L)=0.
\eea
In this model, there are no additional gauge or obvious discrete symmetries carried by the
Standard Model particles, so that the dangerous dimension four proton decay operators
are not necessarily vanishing. 
We refer to Table 6 in Appendix B for a couple of three-generation examples we have found in this setup.
\subsection{Gauge coupling unification}
\label{gaugeunif}

Let us discuss the issue of gauge coupling unification for these models in more detail.
The discussion for the flipped $SU(5)$ models is very similar.
Recall that if one breaks a stringy $SU(5)$ or $SO(10)$ GUT model via discrete
Wilson lines, the Standard Model gauge couplings unify at the GUT scale $2\cdot 10^{16}$GeV, i.e.
at this scale they satisfy the relation $\alpha_3=\alpha_2={5\over 3}\alpha_Y=\alpha_{GUT}$. 
For the weakly coupled heterotic string, however, the Planck scale comes out too low, which
can be remedied in the strong coupling Horava-Witten theory \cite{Horava:1995qa,Witten:1996mz,Horava:1996ma}.
Here it turned out that for the resulting values of $M_{11}$, $\rho$ and $r_{CY}=M_{GUT}^{-1}$, 
the higher order corrections to the gauge couplings could just be ignored compared
to the leading order contributions. 

In our models we expect a completely different picture, as the final $U(1)_Y$ gauge symmetry
has its origin in both $E_8$ walls. 
For the non-abelian gauge couplings of the $SU(3)$ and $SU(2)$ factors including the one-loop contribution we get
\bea
     {1\over  \alpha_{3,2}} &=&{1\over 3\ell_s^6 g_s^2}\int_X J\wedge J\wedge J
              - {1\over  \ell_s^2}
              \int_X J\wedge \left[ -c_2(V)+c_1^2(L)+{1\over 2} c_2(T)\right] \nonumber \\
           &&   + {1\over  \ell_s^2} \sum_a N_a\, \left( {1\over 2} - 
                 \lambda_a \right)^2\,
                 \int_{\Gamma_a} J .
\eea    
Using
\bea
      \eta_{1,1}=60,\quad \eta_{2,2}=4,\quad \kappa_{1,1}=12,\quad \kappa_{2,2}=-4,
\eea 
for the abelian gauge coupling we eventually obtain
\bea
\label{unify}
     {1\over  \alpha_Y} &=& {8\over 3} {1\over \alpha_{3,2}} -
        {1\over  \ell_s^2}
              \int_X J\wedge \left[ c_2(V)+4\,c_1^2(L) \right] + \frac{2}{\ell_s^2} \sum_a N_a \lambda_a \int_{\Gamma_a} J.
\eea
Note that these string models do not give rise to the usual GUT tree level relation
$\alpha_{GUT}={5\over 3}\alpha_Y$ but instead $\alpha_{GUT}={8\over 3}\alpha_Y$, and
therefore the tree level gauge couplings do not unify at $M_{GUT}$ (assuming just the
MSSM matter content). This is the phenomenological prize we have to pay for breaking
the gauge symmetry by non-trivial abelian bundles. After performing the same steps, mutatis mutandis, for our flipped $SU(5)$, we find exactly the same relation (\ref{unify}).

It is striking that just the gauge factor which does not unify with the
other two does receive extra one-loop threshold corrections, so that 
in principle these can help to eventually give a unified picture. 
Defining 
\bea
     {1\over  \alpha_Y} &=& {8\over 3} {1\over \alpha_{GUT}} +  \Delta
\eea
we see that the threshold correction must take the value   $\Delta=-{1\over \alpha_{GUT}}\sim -24$ 
meaning
that
\bea
         {1\over \alpha_Y}\biggl\vert_{\rm 1-loop}=-{3\over 8} {1\over \alpha_Y}\biggr\vert_{\rm tree}. 
\eea
For $\alpha_{GUT}=1/24$, such a relation can just be satisfied with $g_s<1$ and ${\mbox{} r_{CY}>\sqrt{\alpha'}}$
for large enough Chern classes of the vector bundles. 
Of course, in the weakly coupled heterotic framework, the Planck scale still comes out too low
and one should better consider Horava-Witten theory, where now the next-to-leading order
corrections to the gauge couplings should be taken into account. 

Alternatively, one can  contemplate that extra light Higgs fields, if present in the non-chiral spectrum, might lead to gauge 
coupling unification at a different scale. However, this scale is necessarily lower
than the usual GUT scale, which introduces problems with fast proton decay
and worsens the mismatch of the Planck scale. For instance, with 
three Higgs fields the three gauge couplings unify at $M_{GUT}=3 \cdot 10^{11}$ GeV
with a small threshold correction of only $\Delta\sim 2$.

For the example given in section \ref{example} and setting all $\lambda_a =0$ for simplicity, the threshold correction is
\bea
\Delta=-{1\over  \ell_s^2}\int_X J\wedge \left[ c_2(V)+4\,c_1^2(L) \right] = 183r_\sigma-26\rho
\eea
and has the correct sign if $r_\sigma < \frac{26}{183}\rho$. Note that with this choice for $r_\sigma$, the positivity of the gauge couplings can still be achieved.

\section{Conclusions}
The use of non-trivial line bundles considerably enlarges the model building flexibility in heterotic string compactifications. It is the aim of this publication to exemplify this by explicitly constructing realistic GUT and MSSM vacua from simply-connected Calabi-Yau manifolds carrying specific types of $U(N)$ bundles. In trying to avoid the use of discrete Wilson lines for GUT breaking we have been led to the flipped $SU(5)$ scenario, which naturally arises after embedding an $SU(4) \times U(1)$ bundle into one $E_8$ and a particular line bundle into the second. The GUT breaking down to $SU(3) \times SU(2) \times U(1)_Y$ can be further accomplished by a Higgs field in the ${\bf 10}-{\bf \ov {10}}$. If one employs instead a bundle of structure group $SU(5) \times U(1)$, one can directly achieve the MSSM gauge group. The bundle data have to satisfy a couple of constraints which ensure that the hypercharge is indeed massless and that no exotic matter is present. By analysing the resulting equations on an elliptically fibered Calabi-Yau over dP$_4$, we have given a couple of three-generation solutions to these constraints, together with the one imposed by tadpole cancellation and the loop-corrected DUY equation.

So far, we have only explored  the simplest possible choice of bundles in the hidden sector. More refined constructions of this type are conceivable, where the $U(1)_Y$, being a linear combination of abelian factors from both $E_8$'s, might couple the MSSM matter to the (no more) hidden sector matter. This coupling would therefore be  communicated by the photons as the abelian gauge bosons, whose status within the MSSM gauge bosons would thus be of a very specific type. If the hidden matter sector is sufficiently heavy, this can lead to interesting, possibly even experimentally falsifiable effects within the current phenomenological bounds. It also provides the natural arena to study the effects of gauge-mediated SUSY breaking \cite{Diaconescu:2005pc}.

Along the way, we have seen how the presence of five-branes, required by tadpole cancellation, modifies the usual Green-Schwarz mechanism in that additional anomaly counter terms have to be present for a consistent coupling of the five-branes to the heterotic action. The existence of these terms has been confirmed by a direct M-theory computation. In particular, in the Horava-Witten limit, the abelian background bundle on the $E_8$ orbifold fixed plane generates an open-membrane induced D-term potential for the five-brane if it wraps an internal two-cycle whose pull-back to the ten-dimensional end of the world carries abelian gauge flux. A combination of this novel effect with other known contributions to the F- and D-term potential  might have interesting consequences in cosmological applications (see e.g. \cite{Becker:2004gw} and many more articles for discussions of five-brane potentials of cosmological relevance).

We have seen explicitly that one of the two $U(1)$ factors originally present in the model becomes massive via a Stueckelberg-type axionic coupling. In this respect, the heterotic string with $U(N)$ as opposed to $SU(N)$ bundles exhibits precisely the same features which are by now well-known in Type I and Type II orientifold constructions (consult \cite{Blumenhagen:2005mu} for references in that case). A very remarkable consequence of these massive $U(1)$'s has been pointed out recently in \cite{Ibanez:2006qv}, where it is argued that a background value for the field strength of the Kalb-Ramond form along the three non-compact spatial dimensions  leads to a kind of aether carrying the respective global charge of the broken abelian gauge group, with prospects of an alternative solution of the baryon number asymmetry problem. It would be particularly interesting to investigate the applicability of this promising scenario in concrete models also of the heterotic string with line bundles.

The analysis given in this article has only checked that our models indeed exhibit the {\emph {topological}} properties of the three-generation flipped $SU(5)$ and MSSM respectively. It will be important work to compute the non-chiral part of the spectrum including in particular the number of GUT and electro-weak Higgses. Likewise, an analysis of $\mu$-terms and Yukawa-couplings will be indispensible in order to finally decide about the phenomenological relevance of the concrete solutions, possibly along the lines of \cite{Braun:2005xp,Braun:2006me,Bouchard:2006dn}. 

The aspect which is of prime importance to us is that the explicit construction of similar vacua is not tied to Calabi-Yau backgrounds admitting appropriate Wilson lines. Even among the quite restricted class of elliptically fibered Calabi-Yaus we have only analysed a tiny fraction of geometric and bundle parameters when searching for interesting solutions. Many more comparable vacua can probably be constructed. Whether or not they turn out to completely reproduce the observed particle physics, they do exist as consistent solutions within the vast string landscape. A detailed investigation of the topography of this solution space, e.g. in the spirit of  \cite{Dijkstra:2004cc} for the hidden sector of MSSM-like Gepner model orientifolds, \cite{Blumenhagen:2004xx,Gmeiner:2005vz,Gmeiner:2005nh} for a class of toroidal intersecting braneworlds and recently \cite{Dienes:2006ut} for the free-fermionic corner of heterotic string theory, might be one more modest step towards investigating the status of Standard Model like vacua within the landscape. After all, the goal is none less than to understand what a funny world we live in from the point of view of M-theory.

\vskip 1cm
 {\noindent  {\Large \bf Acknowledgements}}
 \vskip 0.5cm 
We gratefully acknowledge  helpful discussions with Florian Gmeiner, Gabriele Honecker and Dimitrios Tsimpis  and stimulating correspondence with Radu Tatar and Taizan Watari. We are indepted to Gabriele Honecker for many  valuable comments on the manuscript.
 \vskip 2cm

\appendix

\section{Normalization of the Green-Schwarz term from Type I string theory}

Since there exists some confusion in the
literature (including our own papers) about the correct normalization of the 
one-loop Green-Schwarz term
\bea
    S_{GS}=c\, \int B_2\wedge X_8
\eea
for the heterotic string and Horava-Witten theory, we will now
thoroughly derive this term in the S-dual Type I theory.
There is overall agreement that the constant $c$ is of the form
\bea
         c={1\over 3\hat c\, (2\pi)^5\, \alpha'},
\eea
but for the parameter $\hat c$ we found various values $\hat c=1,8,16$.

Let us now present our derivation from the S-dual Type IIB orientifold 
point of view. 
The starting point is the well established 
Type IIB action including the Chern-Simons terms
of a stack of $M$  D9-branes, 
\bea
     S_{IIB}&=&{1\over 2\kappa_{10}^2}\int e^{-2\phi} R -
          {1\over 4\kappa_{10}^2}\int G_3\wedge \star G_3 -
         {1\over 2\, g_Y^2}\int e^{-\phi}\, {\rm tr}_{U(M)}
       \left[F\wedge \star F\right]
         + \nonumber \\
    &&\phantom{aaaaaa}
       \mu_9 \int \sum_n C_{2n+2} \wedge {\rm ch}(i{\cal F})\wedge
              \sqrt{\hat A},
\eea
where $\kappa_{10}^2={1\over 2} (2\pi)^7 (\alpha')^4$,
$\mu_9={1\over (2\pi)^9(\alpha')^5}$, ${1\over g_Y^2}=(2\pi\alpha')^2\mu_9$
and the Chern characters ${\rm ch}_k(i{\cal F})={\ell_s^{2k} \over
   k!\, (2\pi)^k}\, {\rm tr}_{U(M)} F$.
The traces are over the fundamental representation of the $U(M)$
gauge theory living on the D9-branes and $G_3=dC_2$ denotes the
R-R three-form field strength.

Now we are taking the orientifold projection which means that we
devide the entire action by a factor of two and introduce
the $\Omega$ image of the stack of branes, which is a stack of
$M$ D9-branes with gauge field $-F$. Tadpole cancellation fixes $M=16$. The Type I action becomes
\bea
     S_{I}&=&{1\over 4\kappa_{10}^2}\int e^{-2\phi} R -
          {1\over 8\kappa_{10}^2}\int  G_3\wedge \star  G_3 -
         {1\over 2\, g_Y^2}\int e^{-\phi} {\rm tr}_{U(16)}\left[ F\wedge
\star F\right]
         + \nonumber \\
   && \mu_9 \int \sum_n  C_{4n+2} \wedge {\rm ch}(i{\cal F})\wedge
              \sqrt{\hat A}.
\eea
Utilizing the trace identities
\bea
             {\rm tr}_{U(16)}[F^2]={1\over 2}{\rm tr}_{SO(32)}[F^2]\, 
\quad\quad
             {\rm tr}_{U(16)}[F^4]={1\over 48}{\rm Tr}_{SO(32)}[F^4],
\eea
with Tr$_{SO(32)}$ denoting the trace in the adjoint representation,
one can write the relevant Chern-Simons terms\footnote{Since we are only
interested in the normalization factors, we just consider the
F-dependent pieces and do not explicitly write down the curvature pieces
arising from the $\hat A$-genus.} 
as
\bea
     S_{I}&=&{1\over 4\kappa_{10}^2}\int e^{-2\phi} R -
          {1\over 8\kappa_{10}^2}\int  G_3\wedge \star  G_3 -
         {1\over 4\, g_Y^2}\int e^{-\phi}\, {\rm tr}_{SO(32)}\left[ F\wedge
\star F\right]
         + \nonumber \\
   && {2\over 4\kappa_{10}^2}{\alpha'\over 4}
      \int C_{6} \wedge {\rm tr}_{SO(32)}[F^2] +
      {1\over 48\, (2\pi)^5 \, \alpha'}\int C_2\wedge
      {1\over 24}\,{\rm Tr}_{SO(32)}[F^4],
\eea
from which one might conclude that the normalization is $\hat c=16$.
This is however not correct as the kinetic terms are not yet canonically
normalized. In order to bring all three kinetic terms to the canonical
form, one has to rescale
\bea
        C_2\to 2\sqrt{2}\, C_2, \quad \alpha'\to \sqrt{2}\,\alpha', \quad
        e^{\phi}\to {1\over 2\sqrt{2}}\, e^{\phi}.
\eea
This leads to the action
\bea
     S_{I}&=&{1\over 2\kappa_{10}^2}\int e^{-2\phi} R -
          {1\over 4\kappa_{10}^2}\int  G_3\wedge \star  G_3 -
         {1\over 2\, \ov g_Y^2}\int e^{-\phi}\, {\rm tr}_{SO(32)}\left[
F\wedge \star F\right]
         + \nonumber \\
   && {2\over 4\kappa_{10}^2}{\alpha'\over 4}
      \int C_{6} \wedge {\rm tr}_{SO(32)}[F^2] +
      {1\over 24\, (2\pi)^5 \, \alpha'}\int C_2\wedge
      {1\over 24}\, {\rm Tr}_{SO(32)}[F^4]
\eea
with the Type I gauge coupling ${1\over \ov g_Y^2}={1\over 2(2\pi)^7 (\alpha')^3}$.
This action is really S-dual to the heterotic string action as given e.g. in \cite{Polchinski:1998rr}.
Therefore we conclude that the correct normalization of the
heterotic Green-Schwarz term is $\hat c=8$, so that
\bea
    S_{GS}= {1\over 24 (2\pi)^5\, \alpha'} \int B_2\wedge  X_8
\eea
if the all the fields are canonically normalized as in \cite{Polchinski:1998rr}.

\section{Three-generation examples}
We list all examples we have found by a computer search on elliptically fibered Calabi-Yau spaces with base spaces $\dP_r$, $r=1,\ldots,4$ and the Hirzebruch surfaces $F_r$ in a range from $-10,\ldots,10$ for all parameters. We have found three-generation models only on $\dP_4$. Table 5 contains the three-generation flipped $SU(5)$ models, whereas in table 6 we list all three-generation vacua directly with MSSM gauge group (see section 5) which we have found.

\begin{table}[htb]
\renewcommand{\arraystretch}{1.5}
\begin{center}
\begin{scriptsize}
\begin{tabular}{|c||c|c|c|c|}
\hline
\hline
$\lambda$ & $\eta$ & $q$ & $c_1(\zeta)$ & $[W]$ \\
\hline \hline
$\frac{1}{4}$ & $14l - 2E_1 - 6E_2 - 6E_3 - 2E_4$ & $0$ & $-l+E_2+E_3+E_4$ & $27F+(22l-10E_1-6E_2-6E_3-10E_4)\sigma$\\
\hline
$\frac{1}{4}$ & $18l - 10E_1 - 6E_2 - 6E_3 - 6E_4$ & $0$ & $-l+E_2+E_3+E_4$ & $27F+(18l-2E_1-6E_2-6E_3-6E_4)\sigma$\\
\hline
$\frac{1}{4}$ & $14l - 6E_1 - 2E_2 - 2E_3 - 6E_4$ & $0$ & $-E_1+E_4$ & $27F+(22l-6E_1-10E_2-10E_3-6E_4)\sigma$\\
\hline
$\frac{1}{4}$ & $14l - 2E_1 - 6E_2 - 6E_3 - 2E_4$ & $0$ & $-E_1+E_4$ & $27F+(22l-10E_1-6E_2-6E_3-10E_4)\sigma$\\
\hline
$\frac{1}{4}$ & $18l - 6E_1 - 10E_2 - 6E_3 - 6E_4$ & $0$ & $-E_1+E_4$ & $27F+(18l-6E_1-2E_2-6E_3-6E_4)\sigma$\\
\hline
$\frac{1}{4}$ & $14l - 2E_1 - 6E_2 - 6E_3 - 2E_4$ & $0$ & $l-E_1-E_2-E_3$ & $27F+(22l-10E_1-6E_2-6E_3-10E_4)\sigma$\\
\hline
$\frac{1}{4}$ & $18l - 6E_1 - 6E_2 - 6E_3 - 10E_4$ & $0$ & $l-E_1-E_2-E_3$ & $27F+(18l-6E_1-6E_2-6E_3-2E_4)\sigma$\\
\hline
\end{tabular}
\end{scriptsize}
\caption{Flipped $SU(5)\times U(1)_X$ models on $\dP_4$.}
\end{center}
\end{table}

\begin{table}[htb]
\renewcommand{\arraystretch}{1.5}
\begin{center}
\begin{scriptsize}
\begin{tabular}{|c||c|c|c|c|}
\hline
\hline
$\lambda$ & $\eta$ & $q$ & $c_1(\zeta)$ & $[W]$ \\
\hline \hline
$\frac{1}{2}$ & $15l-3E_1-5E_2-5E_3-5E_4$ & $0$ & $-l+E_2+E_3+E_4$ & $7F+(21l-9E_1-7E_2-7E_3-7E_4)\sigma$\\
\hline
$\frac{1}{2}$ & $15l-2E_1-5E_2-5E_3-5E_4$ & $0$ & $-l+E_2+E_3+E_4$ & $7F+(21l-10E_1-7E_2-7E_3-7E_4)\sigma$\\
\hline
$\frac{1}{2}$ & $17l-7E_1-7E_2-5E_3-5E_4$ & $0$ & $-l+E_2+E_3+E_4$ & $7F+(19l-5E_1-5E_2-7E_3-7E_4)\sigma$\\
\hline
$\frac{1}{2}$ & $18l-8E_1-8E_2-5E_3-5E_4$ & $0$ & $-l+E_2+E_3+E_4$ & $7F+(18l-4E_1-4E_2-7E_3-7E_4)\sigma$\\
\hline
$\frac{1}{2}$ & $20l-3E_1-10E_2-10E_3$ & $0$ & $-l+E_2+E_3+E_4$ & $7F+(16l-9E_1-2E_2-2E_3-12E_4)\sigma$\\
\hline
$\frac{1}{2}$ & $20l-2E_1-10E_2-10E_3$ & $0$ & $-l+E_2+E_3+E_4$ & $7F+(16l-10E_1-2E_2-2E_3-12E_4)\sigma$\\
\hline
$\frac{1}{2}$ & $15l-5E_1-5E_2-3E_3-5E_4$ & $0$ & $-E_1+E_4$ & $7F+(21l-7E_1-7E_2-9E_3-7E_4)\sigma$\\
\hline
$\frac{1}{2}$ & $15l-5E_1-5E_2-2E_3-5E_4$ & $0$ & $-E_1+E_4$ & $7F+(21l-7E_1-7E_2-10E_3-7E_4)\sigma$\\
\hline
$\frac{1}{2}$ & $15l-5E_1-3E_2-5E_4$ & $0$ & $-E_1+E_4$ & $7F+(21l-7E_1-9E_2-12E_3-7E_4)\sigma$\\
\hline
$\frac{1}{2}$ & $15l-5E_1-2E_2-5E_4$ & $0$ & $-E_1+E_4$ & $7F+(21l-7E_1-10E_2-12E_3-7E_4)\sigma$\\
\hline
$\frac{1}{2}$ & $15l-5E_2-3E_3$ & $0$ & $-E_1+E_4$ & $7F+(21l-12E_1-7E_2-9E_3-12E_4)\sigma$\\
\hline
$\frac{1}{2}$ & $17l-7E_1-5E_2-5E_3-7E_4$ & $0$ & $-E_1+E_4$ & $7F+(19l-5E_1-7E_2-7E_3-5E_4)\sigma$\\
\hline
$\frac{1}{2}$ & $17l-7E_1-5E_2-7E_4$ & $0$ & $-E_1+E_4$ & $7F+(19l-5E_1-7E_2-12E_3-5E_4)\sigma$\\
\hline
$\frac{1}{2}$ & $17l-7E_1-7E_4$ & $0$ & $-E_1+E_4$ & $7F+(19l-5E_1-12E_2-12E_3-5E_4)\sigma$\\
\hline
$\frac{1}{2}$ & $17l-5E_1-7E_2-7E_3-5E_4$ & $0$ & $-E_1+E_4$ & $7F+(19l-7E_1-5E_2-5E_3-7E_4)\sigma$\\
\hline
$\frac{1}{2}$ & $17l-7E_2-7E_3$ & $0$ & $-E_1+E_4$ & $7F+(19l-12E_1-5E_2-5E_3-12E_4)\sigma$\\
\hline
$\frac{1}{2}$ & $18l-8E_1-5E_2-5E_3-8E_4$ & $0$ & $-E_1+E_4$ & $7F+(18l-4E_1-7E_2-7E_3-4E_4)\sigma$\\
\hline
$\frac{1}{2}$ & $18l-8E_1-5E_2-8E_4$ & $0$ & $-E_1+E_4$ & $7F+(18l-4E_1-7E_2-12E_3-4E_4)\sigma$\\
\hline
$\frac{1}{2}$ & $18l-8E_1-8E_4$ & $0$ & $-E_1+E_4$ & $7F+(18l-4E_1-12E_2-12E_3-4E_4)\sigma$\\
\hline
$\frac{1}{2}$ & $18l-5E_1-8E_2-8E_3-5E_4$ & $0$ & $-E_1+E_4$ & $7F+(18l-7E_1-4E_2-4E_3-7E_4)\sigma$\\
\hline
$\frac{1}{2}$ & $18l-8E_2-8E_3$ & $0$ & $-E_1+E_4$ & $7F+(18l-12E_1-4E_2-4E_3-12E_4)\sigma$\\
\hline
$\frac{1}{2}$ & $20l-10E_1-5E_2-3E_3-10E_4$ & $0$ & $-E_1+E_4$ & $7F+(16l-2E_1-7E_2-9E_3-2E_4)\sigma$\\
\hline
$\frac{1}{2}$ & $20l-10E_1-5E_2-2E_3-10E_4$ & $0$ & $-E_1+E_4$ & $7F+(16l-2E_1-7E_2-10E_3-2E_4)\sigma$\\
\hline
$\frac{1}{2}$ & $20l-10E_1-3E_2-10E_4$ & $0$ & $-E_1+E_4$ & $7F+(16l-2E_1-9E_2-12E_3-2E_4)\sigma$\\
\hline
$\frac{1}{2}$ & $20l-10E_1-2E_2-10E_4$ & $0$ & $-E_1+E_4$ & $7F+(16l-2E_1-10E_2-12E_3-2E_4)\sigma$\\
\hline
$\frac{1}{2}$ & $15l-5E_1-5E_2-5E_3-3E_4$ & $0$ & $l-E_1-E_2-E_3$ & $7F+(21l-7E_1-7E_2-7E_3-9E_4)\sigma$\\
\hline
$\frac{1}{2}$ & $15l-5E_1-5E_2-5E_3-2E_4$ & $0$ & $l-E_1-E_2-E_3$ & $7F+(21l-7E_1-7E_2-7E_3-10E_4)\sigma$\\
\hline
$\frac{1}{2}$ & $17l-7E_1-5E_2-5E_3-7E_4$ & $0$ & $l-E_1-E_2-E_3$ & $7F+(19l-5E_1-7E_2-7E_3-5E_4)\sigma$\\
\hline
$\frac{1}{2}$ & $18l-8E_1-5E_2-5E_3-8E_4$ & $0$ & $l-E_1-E_2-E_3$ & $7F+(18l-4E_1-7E_2-7E_3-4E_4)\sigma$\\
\hline
$\frac{1}{2}$ & $20l-10E_1-10E_2-3E_4$ & $0$ & $l-E_1-E_2-E_3$ & $7F+(16l-2E_1-2E_2-12E_3-9E_4)\sigma$\\
\hline
$\frac{1}{2}$ & $20l-10E_1-10E_2-2E_4$ & $0$ & $l-E_1-E_2-E_3$ & $7F+(16l-2E_1-2E_2-12E_3-10E_4)\sigma$\\
\hline
\end{tabular}
\end{scriptsize}
\caption{$SU(3) \times SU(2) \times U(1)$ models on $\dP_4$.}
\end{center}
\end{table}

\clearpage
\nocite{*}
\bibliography{rev}
\bibliographystyle{utphys}

\end{document}